\documentstyle[preprint,amstex,prbbib,epsf,aps]{revtex}

\begin{document}
\draft  
\title{Correlation function spectroscopy of inelastic
lifetime in heavily doped GaAs}

\author{J. K\"onemann,$^1$ P. K\"onig,$^1$ T. Schmidt,$^1$
 E. McCann,$^2$ Vladimir I. Fal'ko,$^2$ and R. J. Haug$^1$}

\address{$^1$ Institut f\"ur Festk\"orperphysik, Universit\"at
Hannover, Appelstr. 2, 30167 Hannover, Germany\\
$^2$ Department of Physics, Lancaster University,
Lancaster, LA1 4YB, United Kingdom}

\date{\today}
\maketitle
\begin{abstract}
Measurements of resonant tunneling through
a localized impurity state are used to probe
fluctuations in the local density of states of heavily doped GaAs.
The measured differential conductance is analyzed in terms 
of correlation functions with respect to voltage.
A qualitative picture based on the scaling theory
of Thouless is developed to relate the observed fluctuations
to the statistics of single particle wavefunctions. In
a quantitative theory correlation functions are calculated.
By comparing the experimental and theoretical correlation functions
the effective dimensionality of the emitter is analyzed and
the dependence of the inelastic lifetime on energy is extracted.

\end{abstract}
\vspace*{2em}   

\pacs{PACS numbers: 73.23.-b, 72.20.My, 85.30.Mn}

\narrowtext

\bibliographystyle{simpl1}

\section{Introduction}
%
The observation of impurity-assisted tunneling in vertical transport 
experiments on double-barrier semiconductor
heterostructures\cite{su92,del92,tew92,des96,sch96} led to the possibility
of using the resonant impurity level as a local probe of
electronic states of electrodes prepared from heavily doped degenerate 
semiconductors.
A number of experiments in strongly asymmetric double-barrier
structures have measured directly the local density of states (LDOS) of an
electrode as a function of excitation energy $E$
from the Fermi level,\cite{sch96,sch97,mai00,hol00,kon2000a,kon2001} 
those with the highest spectral resolution 
reporting features including Zeeman splitting of single-particle levels 
in a disordered emitter.\cite{kon2000a}

\begin{figure}[t]
%
\centerline{\epsfxsize=\hsize
\epsffile{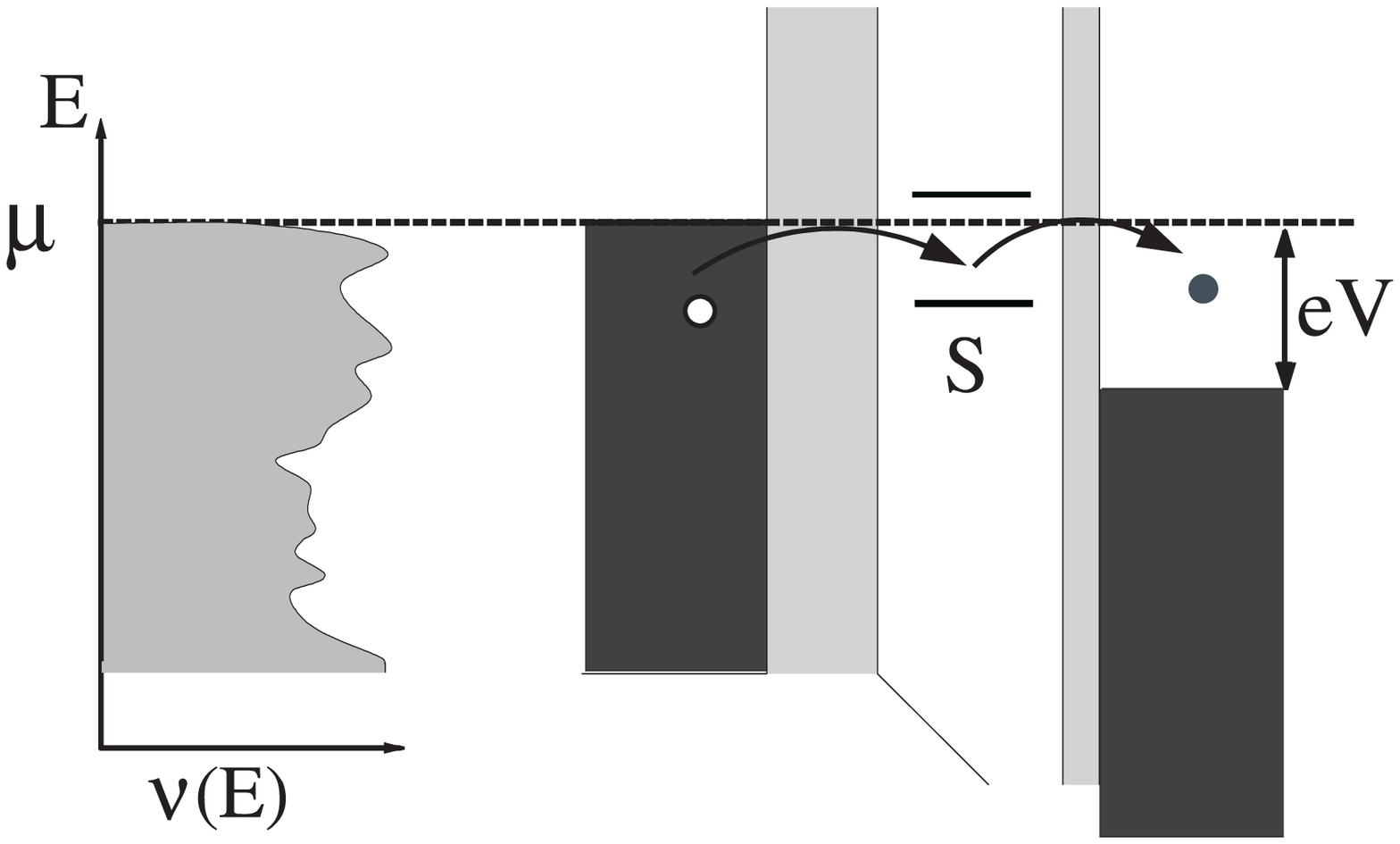}}
\refstepcounter{figure}
\label{fig:idea}
{\setlength{\baselineskip}{10pt} FIG.\ \ref{fig:idea}.
Sketch of the resonant tunneling spectroscopy of the LDOS using an 
impurity state in a double-barrier structure.
Electrons tunnel from a heavily-doped disordered emitter
through the energetically-lowest level $S$ of the quantum well
sandwiched between the double barriers,
so that $S$ serves as a spectrometer of the density of states $\nu (E)$
of the emitter.}
\end{figure}

The idea of such experiments is illustrated by the sketch in 
Fig.~\ref{fig:idea}.
Electrons tunnel from a heavily-doped disordered emitter
through the energetically lowest level of the quantum well
sandwiched between the double barriers. This energetically lowest level of 
the quantum well 
serves as the spectrometer $S$.
At zero bias, the energy of this impurity level, $E_S$, does not
coincide with the chemical potential $\mu$ in the emitter. It comes to
resonance only after the bias voltage reaches a threshold value $V_S(E_S)$.
Typical current-voltage I(V) characteristics of such a device can be divided into 
three intervals:\cite{del92,tew92,des96,sch96,Sivan} 
one interval below the threshold, where $I\approx 0$;  the threshold regime 
$V\approx V_S(E_S)$, where $I(V)$ undergoes a jump when 
the resonant level crosses the Fermi level $\mu$ in the emitter, and 
the interval of a plateau, $V_S(E_S)<V<V_1(E_1)$, where the current 
remains nearly constant. The latter interval lasts until the next 
impurity level $E_1$ is lowered enough to contribute to the transport 
and it is ideal for studying the image of the LDOS 
in the emitting reservoir,\cite{sch96,fal97} since any further variation of 
the current as
a function of bias voltage, $I(V)$ is dominated by the energy 
dependence of the tunneling density of states in the emitter, 
$I(V)\propto\nu (E)$. A convenient way to look at such I(V) characteristics is
to plot  the differential conductance 
$G(V) = dI/dV \propto d\nu/dE$, in which the image of variation in LDOS is 
more pronounced.
   


In a disordered medium, the energy dependence of the LDOS studied at a certain 
point of a sample reveals an irregular fine structure,\cite{Lerner}  
$\nu(E)=\nu_0 +\delta\nu (E)$,
which arises from quantum interference of elastically scattered quasi-particles
diffusing coherently within a length scale related to their 
lifetime at a particular energy. Since the Aharonov-Bohm phase accumulated by 
a diffusing particle in a magnetic field changes the interference pattern, such 
a fine structure, $\delta\nu (E)$ depends randomly on a magnetic field $B$.
In tunneling experiments, 
the interfering quasiparticle is, in fact, a `hole' in the Fermi sea left behind by 
the tunneling of an electron out of the emitter with $E<\mu$. Being in a 
non-equilibrium state, such a hole `floats up' towards the Fermi level, due to 
inelastic collisions between electrons, so that it can be characterized by 
a finite lifetime equivalent to a broadening of emitter states.\cite{Sivan} 
The broadening of emitter states washes out 
the finest features in the LDOS fluctuations, and, therefore, it strongly affects
the amplitude and correlation parameters of fluctuations of the differential 
conductance\cite{fal97} of a given resonant tunneling device, 
$\delta G(V) = G(V) - \langle G \rangle$. A particularly convenient situation
to study fluctuations is realized in devices where the mean value of the 
density of states in the emitter and also the transmission through the barriers
varies much slower than fine fluctuations in LDOS, so that 
within the narrow energy interval below the Fermi energy of the emitter 
$\langle G \rangle$ is negligible and $\delta G(V)\approx G(V)$. 

Recently, we reported\cite{kon2001} an experiment where the speeding up of the
quasi-particle relaxation upon the increase of the energetic distance 
to the Fermi level (equal to the excitation energy of the Fermi sea holes) 
was observed via the decline in the variance of differential conductance 
fluctuations at higher bias voltages, $\langle (\delta G)^2 \rangle$. 
In the present publication, we study the 
correlation function, $K(V)$, of a random differential conductance pattern worked 
out for different bias voltage intervals. To make this analysis 
sound, the differential conductance of a resonant tunneling structure 
has been measured for a dense grid of magnetic field values, which has largely 
increased the statistical ensemble of data used in the 
evaluation of the correlation function $K(\Delta V)$ of a random pattern $\delta G(V,B)$,
\begin{eqnarray}\label{e11}
 K(\Delta V) = 
{{ \langle \delta G(V + \Delta V ,B)\delta G(V,B)\rangle_B} \over
{\langle (\delta G)^2 \rangle}}, 
\end{eqnarray}
and has allowed us to compare details of its shape to the results
of a theoretical analysis. The latter effort has enabled us to notice some
geometrical features of the structure used in this experiment (produced in
a particular growth process), which would be difficult to detect otherwise. 

The material presented in this paper is organized as follows.
In Section II, the experimental set-up, parameters 
and design of the structures which we used, and the raw 
data of G(V) characteristics are discussed.
In order to characterize the spectral resolution of the spectrometer, we 
analyze in Section II the form of $\langle G(V)\rangle_B$ averaged over 
many runs taken at different values of applied magnetic field.
A detailed quantitative analysis of fluctuations and their correlation 
functions is presented in Section III, in comparison to the
results of a theory presented in two Appendices.  The end of Section III is 
devoted to the discussion of energy dependence of the quasi-particle 
relaxation rate extracted from this analysis, from the point of view
of Aronov-Altshuler theory of electron-electron interaction in disordered metals\cite{A+A79,siv94}. 
Appendix A completes the text with a qualitative estimation of 
the variance of the differential conductance fluctuations based upon the theory 
of statistical and correlation properties of chaotic wavefunctions in disordered media
using an approach similar to Thouless's scaling theory.\cite{tho77} 
The quantitative analysis of the variance and correlation properties of 
a pattern of 
$\delta G(V)$, including the dimensional crossover, is presented in Appendix B.

\section{Samples and experimental results \label{SEx}} 
\begin{figure}
\epsfxsize=0.7\hsize
\epsffile{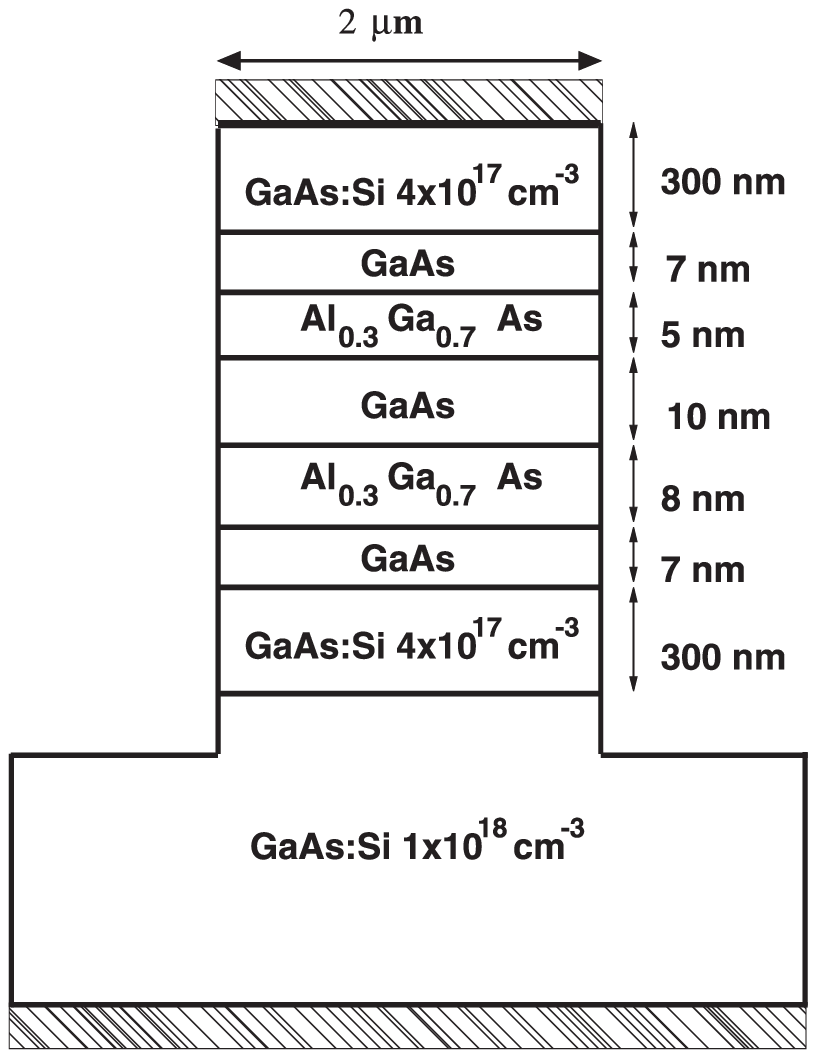}
\refstepcounter{figure}
\label{fig:het}
{\setlength{\baselineskip}{10pt} FIG.\ \ref{fig:het}.
Layer structure of the double-barrier heterostructure.}
\end{figure}

The experiment was performed using an asymmetric
double-barrier heterostructure which was grown by molecular-beam epitaxy on 
$n^+$-type GaAs substrate. Directly on top of the substrate the layer sequence
for the resonant tunneling diode was grown as illustrated in Fig.~\ref{fig:het}. 
The growth started with a 300~nm thick GaAs layer doped with Si to $4.0\times 10^{17}$~cm$^{-3}$. 
This emitter layer is followed by a very thin spacer layer of 7~nm undoped GaAs. 
The actual resonant tunneling structure consists of a 
$10$~nm wide
GaAs quantum  well sandwiched between two Al$_{0.3}$Ga$_{0.7}$As
barriers of $5$ and $8$~nm width (top and bottom barrier).
The collector of the structure is formed by a second spacer layer of 7~nm undoped GaAs 
and a 300~nm thick layer of  GaAs doped with Si to $4.0\times 10^{17}$~cm$^{-3}$.

The barrier structure
is highly asymmetric, the transparency of the thick emitter
barrier is much lower than that of the collector
barrier, which means that 
the value of the tunneling current is dominated by the low transmission of
the emitter barrier. Due to the thin spacer layer the nominally undoped quantum well 
contains a small number of residual impurities. The energetically lowest impurity state
will be used as a local spectrometer of the emitter states.

\begin{figure}[t]
%
\epsfxsize=\hsize
\epsffile{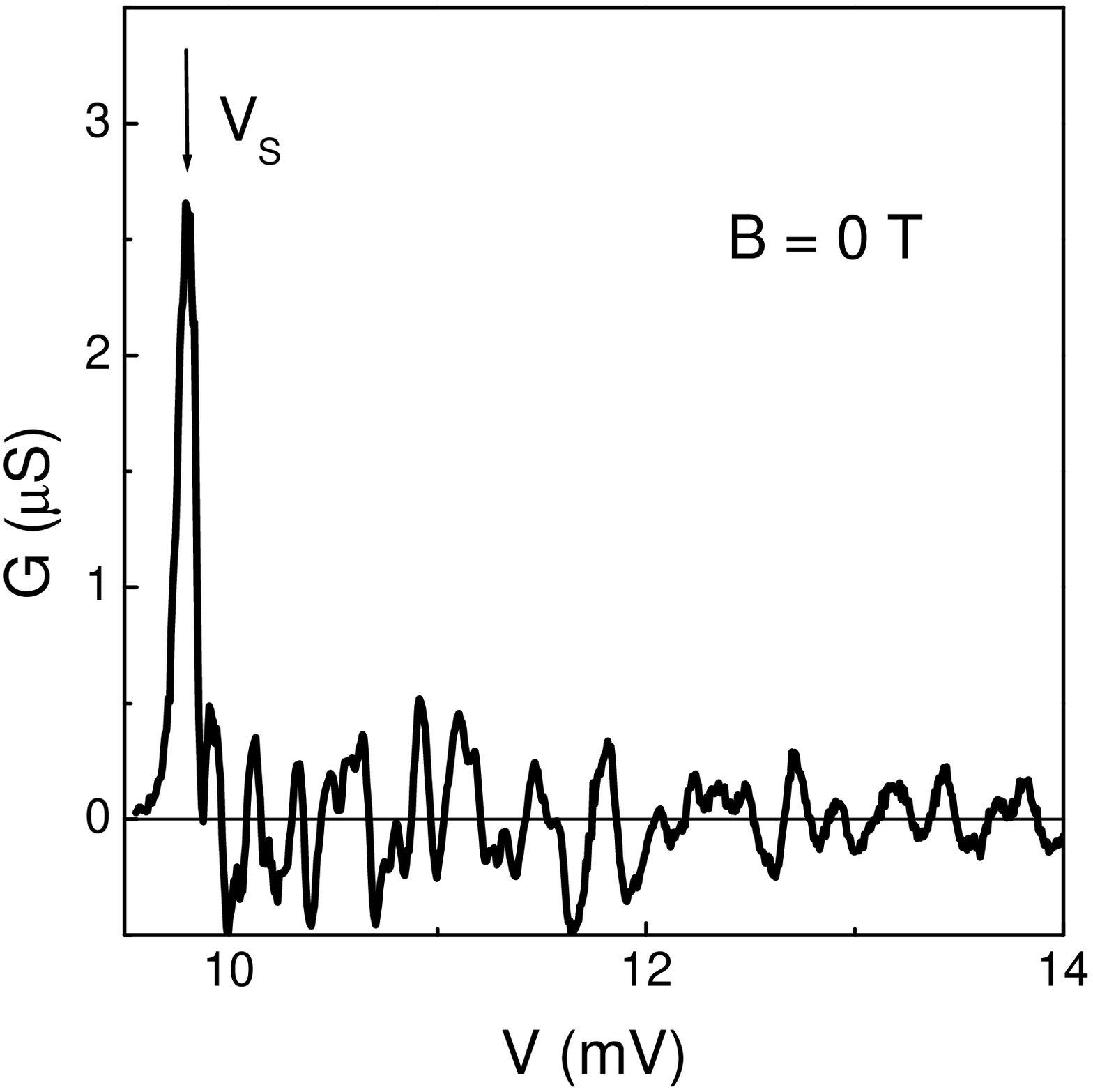}
\refstepcounter{figure}
\label{fig:gv}
{\setlength{\baselineskip}{10pt} FIG.\ \ref{fig:gv}.
Image of LDOS fluctuations: Typical plot of the differential 
conductance $G$ versus bias voltage $V$ at $B=0$~T
and a base temperature of $T=20$~mK.} 
\end{figure}

In order to limit the number of residual impurities in the quantum well, 
pillars with small areas were fabricated from this heterostructure.\cite{fab_tew} 
By employing electron-beam lithography, evaporation and lift-off, 
AuGe/Ni layers were deposited  on the top of the wafer. This
metalization served both as an ohmic contact 
and as an etch mask for the following reactive ion etching (RIE) step. 
A AuGe/Ni coating was also evaporated onto the substrate side of the wafer 
to form the back ohmic contact. Free standing pillars 
with diameters in the $\mu$m and sub-$\mu$m range and a typical height of 
several hundred nm were etched using RIE. Then, large-area Cr/Ag/Au 
bond pads could be prepared on top of the pillars by planarizing 
the pillars with an insulating polyimide layer.
%
The tunneling current 
was measured with a dc technique in a dilution refrigerator at 20~mK base temperature. 
For our analysis the differential conductance $G(V)$ was numerically calculated
from the measured current values. A typical $G(V)$ trace 
is shown in Fig.~\ref{fig:gv}.
At zero bias, $S$ lies above the Fermi level in the emitter 
and is not available for resonant transport, resulting in $G=0$. 
At $V_S=9.8$~mV, the spectrometer crosses the Fermi level and the 
current jumps abruptly from zero to a finite value, 
resulting in a sharp peak in the derivative $G \propto d\nu/d\epsilon$.
For larger bias voltages a reproducible oscillatory 
fine structure can be seen, which we attribute as 
the result of LDOS fluctuations in the contact regions.
This fine structure is formed by electrons which tunnel from 
below the emitter Fermi level through the lowest discrete state
$S$ in the quantum well.  Since the emitter barrier is 
much stronger than the collector one, the value of the current 
step is mainly determined by the tunneling rate $\Gamma_l/\hbar$ 
through the thick barrier on the emitter side.
Due to this large barrier asymmetry the 
G(V) curve at voltages $V > 9.8$~mV  represents the energy dependence
only of the LDOS $d\nu/d\epsilon$
 in the emitter contact.  
So the fine structure represents an image of emitter contact 
LDOS fluctuations scanned
by the impurity-related level in the quantum well.\cite{sch96} 

\begin{figure}
%
\epsfxsize=0.8\hsize
\epsffile{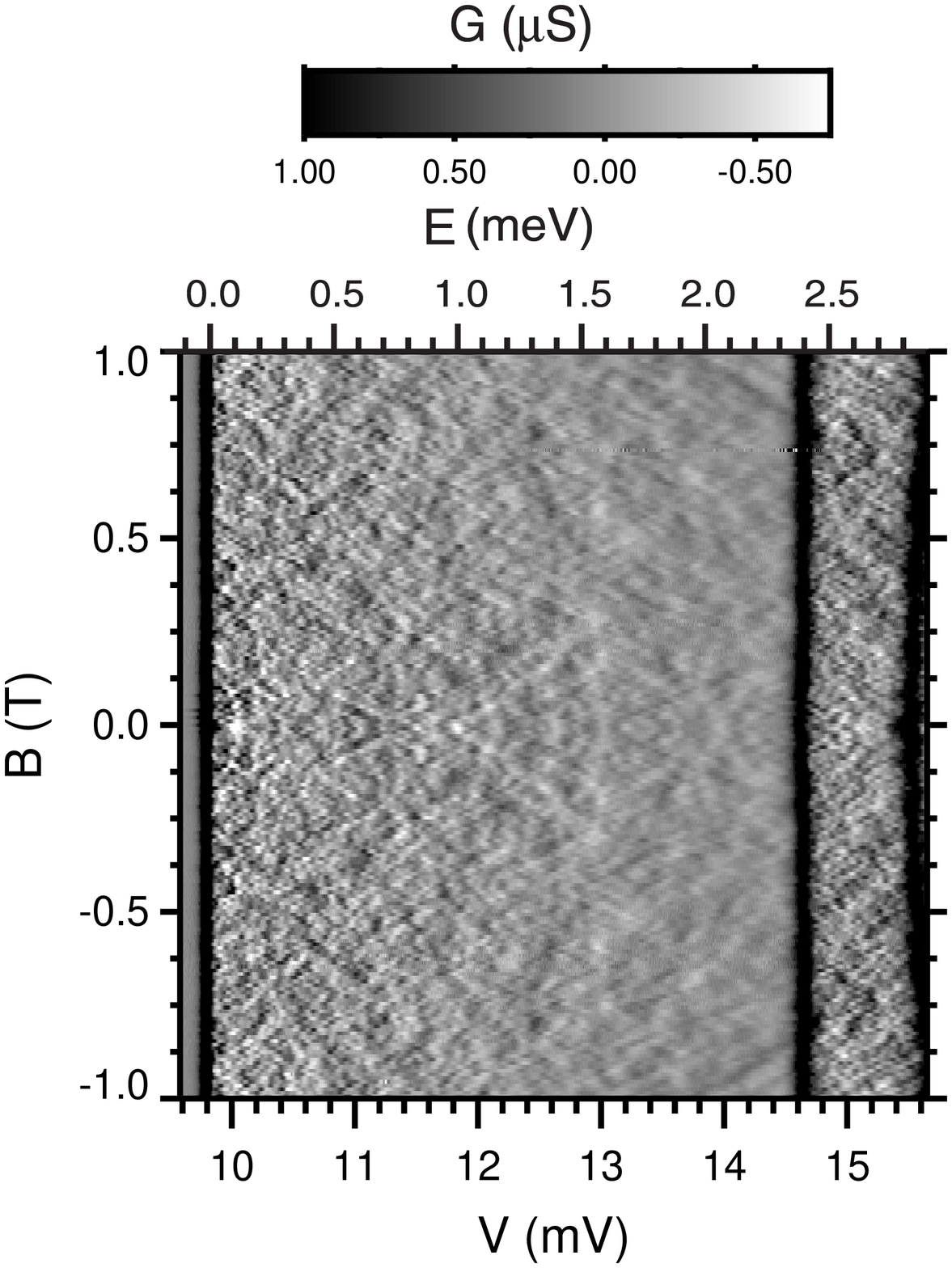}
\refstepcounter{figure}
\label{fig:map}
{\setlength{\baselineskip}{10pt} FIG.\ \ref{fig:map}.
Contour plot of the differential conductance $G$ as a function of bias voltage $V$ (step $7$~$\mu$V) and magnetic field $B\parallel I$ 
(step $10$~mT) for $T=20 $~mK. The excitation energy $E$ on the top scale is converted from the bias voltage $V$, see text.} 
\end{figure}

The quantum interference interpretation of the observed fine 
structure is supported by the observed effect of an applied magnetic field.
The oscillatory form of $G(V)$ randomly changes upon variation of a magnetic field,
at the scale $\Delta B<30$~mT. 
Figure~\ref{fig:map} shows a grey-scale 
image of the differential conductance measured as a function of both bias 
voltage and a magnetic field within the interval of
fields $-1$~T $<B<1$~T, where the Landau quantization of states in the emitter 
is completely suppressed by disorder. This diagram is symmetric 
with respect to magnetic field inversion, as it should be 
for a two-terminal measurement. The use of magnetic field 
enables us to get a sound amount of data for the following statistical 
analysis of fluctuations.  

%

The onset of resonant tunneling through the lowest lying impurity state 
$S$ appears in Fig.~\ref{fig:map} as a black line at a voltage of 9.8~mV (in 
parallel to the B-axis). The second black line 
at a bias voltage of 14.6~mV appears when the next, higher lying
impurity state crosses the Fermi energy in the emitter. 
For voltages ranging from 9.8~mV up to 14.6~mV, the measured tunneling current
results only from tunneling through the lowest lying impurity state $S$.
This state is used as a local spectrometer to scan the LDOS below
the Fermi level in the emitter.
Voltage and energy scales are related via 
$E=\alpha e (V-V_S)$ where the prefactor $\alpha=0.5$ accounts for the fact that only part of the 
voltage drops between emitter and spectrometer.\cite{su92,del92,des96} 
Therefore, the plot in  Fig.~\ref{fig:map} covers an energy range of 
quasi-particle excitations of about $0\leq E\leq 2.4$~meV,
which is indicated by an alternative scale for the
horizontal axis on the upper side of this figure. 
The amplitude of $G(V)$ fluctuations decreases (fine structure washes out) with 
increasing bias voltage in the range
$9.8$~mV $< V < 14.6$~mV (interval between two peaks).
At the same time, the characteristic voltage scale dominating the fine structure 
increases which is interpreted below to be the result of the 
inelastic broadening of quasi-particle states in the emitter. 
Note that although oscillations at larger energy scales are also present
in $I(V)\propto \nu (E)$, their contribution to
$G(V)$ is suppressed due to the differentiation.
For a broad spectrometer both the amplitude
and correlation voltage of fluctuations would be the same over
the entire range of $V_S \leq V \leq V_{1}$.
For a narrow spectrometer, as studied in the present work,
inelastic broadening of states in the bulk exceeds the spectrometer
width upon increasing the excitation energy of a quasi-hole
left in the emitter. Then, this inelastic broadening affects the parameter of the fluctuation pattern.
Note that the observed fluctuations become sharp and large again 
after the second impurity level begins to contribute to the current at $V > 14.6$~mV. 
This is because tunneling 
through the second impurity state involves states close to the Fermi energy
which have negligible inelastic broadening.
In the following, we shall focus on the tunneling through 
the lowest lying impurity
state, i.e. on the interval of bias voltages smaller than $V<14.6$~mV.

\begin{figure}
%
\epsfxsize=\hsize
\epsffile{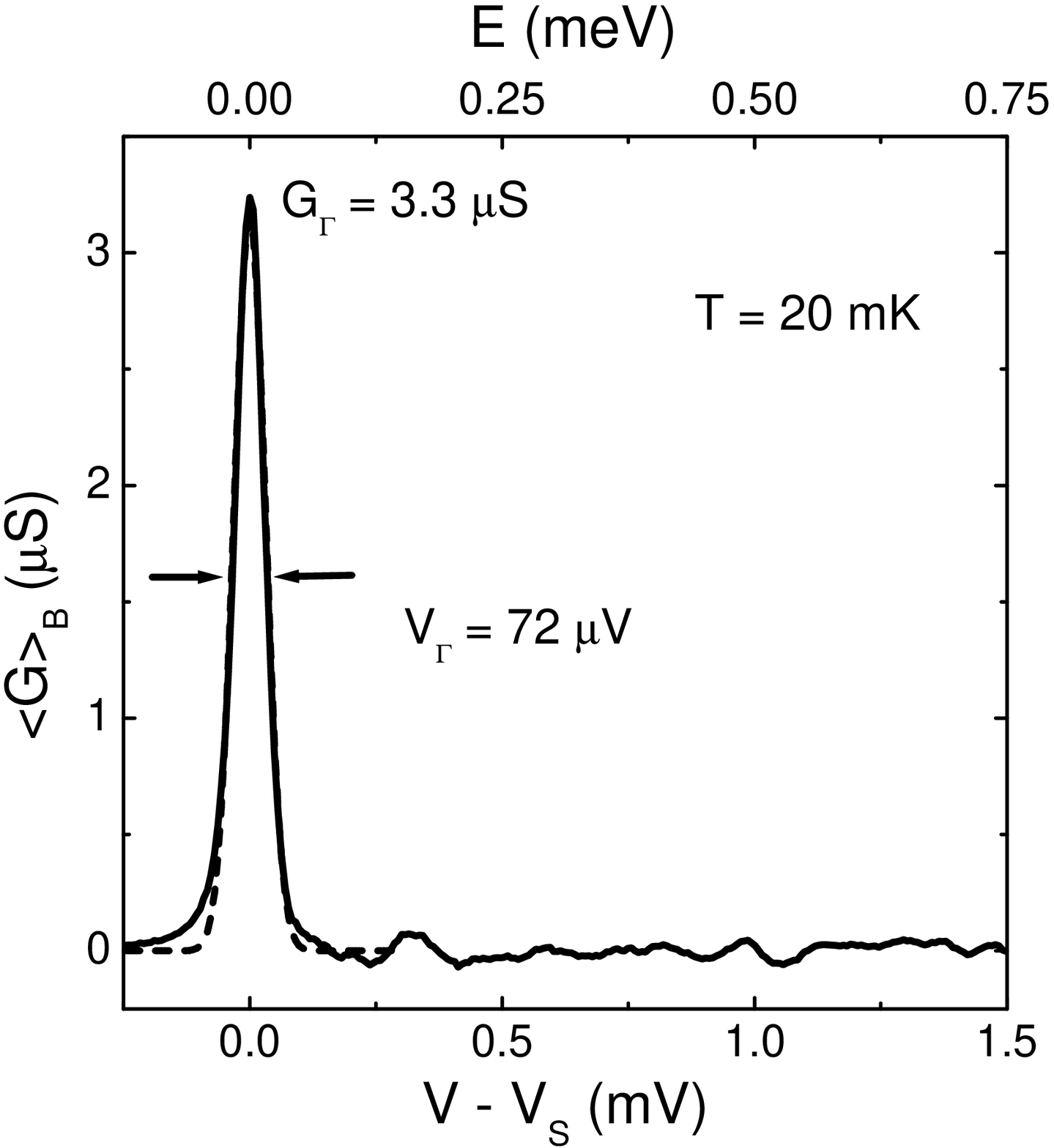}
\refstepcounter{figure}
\label{fig:gv_av}
{\setlength{\baselineskip}{10pt} FIG.\ \ref{fig:gv_av}.
The averaged differential conductance $\left<G(B)\right>_B$ of the device  
obtained as described in the text.} 
\end{figure}

In Fig.~\ref{fig:gv_av} the differential conductance is shown 
after averaging the raw data $G(V,B)$ over the interval 
of magnetic field specified above.
This averaging increases the contrast between the main peak 
corresponding to the spectrometer $S$ crossing the emitter Fermi level 
and $\left<G(V)\right>_B$ at larger bias voltages, where 
a random contribution from LDOS fluctuations is strongly suppressed. The
fluctuations are suppressed by a 
statistical weight 
of $\sqrt{N}$, where $N$ is the number of uncorrelated G(V) traces 
taken at various magnetic fields.
The plot in Fig.~\ref{fig:gv_av} can be used to extract the 
nominal spectrometer width, $\Gamma$.
The emitter barrier in the device we study is 
thicker than the collector one, so that the broadening of the 
resonant level is dominated by electron escape from it to the 
collector, $\Gamma =\Gamma_r+\Gamma_l\approx\Gamma_r$, whereas 
the value of the current step is mainly determined
by the tunneling rate $\Gamma_l/\hbar$ through the thick barrier on the emitter side.  
The averaged $\left<G(V)\right>_B$ characteristics 
at the threshold can be parameterized\cite{Chaplik,Azbel,ResTunn} 
by the height of the conductance peak at 
the threshold voltage $V_S$ and by its width $V_{\Gamma}$ 
at the half-maximum, which is given by,\cite{RemSpin} 
\begin{equation}
V_{\Gamma} \approx \Gamma / (e\alpha ). 
\end{equation}
Below, we use $\Gamma =36$~$\mu$eV taken directly from Fig.~\ref{fig:gv}.

%

\section{Statistical characteristics of the differential
 conductance fluctuations pattern.}

In this section, we analyze correlation properties of the measured 
differential conductance pattern, aiming to extract 
from this the value and the form of the energy dependence of the  
decay rate for quasi-particles. The relevance of the correlation function of the 
fluctuations pattern for such an analysis arises because the auto-correlation function 
of fluctuations reflects the typical scale of their energy dependence (which
is equivalent information to that in the power spectrum of frequencies of 
oscillations).  To start with, the pattern of $\delta G(V)$ 
is random and it is related to the derivative of the LDOS with respect to 
energy, where the contribution from features at the finest energy scale is enhanced by 
differentiation. Therefore, the correlation function 
$K(\Delta V)=\langle \delta G(V+\Delta V,B)\delta G(V,B)\rangle / 
\langle \delta G^2 \rangle$ carries information about the finest 
resolution of quantum states in the emitter. 
On the one hand, due to the finite spectrometer width, $\Gamma$, fine 
structure in the LDOS at energy scales 
smaller than $\Gamma$ is smeared out by the spectrometer, so that in the measurements
reported above it cannot be resolved. On the other hand, the finest energy
scale of LDOS fluctuations is intrinsically limited by inelastic broadening of 
quasi-particle states in the emitter, $\hbar \gamma$. As a result, the 
typical value of bias voltage, at which the differential conductance varies
randomly, is determined by the sum of the above two,
\begin{equation}
V_c = \frac{1}{\alpha} [\Gamma +\hbar\gamma] = E_c/\alpha,
\end{equation}
where the spectrometer width $\Gamma$ is the same for the entire interval of energies
of the quasi hole in the emitter (left behind by the tunneling process) that we
are able to study using one impurity state, whereas the inelastic broadening,
$\hbar\gamma(E)$, is dependent on the excitation energy and varies across 
the studied bias voltage interval.
The same combination of energetic parameters also determines the variance of the 
differential conductance fluctuations, $\langle \delta G^2 \rangle$, which will
be discussed in subsection B.

\subsection{Correlation function of fluctuations.}

The experimental determination of the correlation function, $K$, consists of the 
evaluation of the variance $\langle \delta G^2 \rangle_B$ and, then, the autocorrelation 
function of the measured 
differential conductance fluctuations pattern,
$$K(\Delta V)=\langle \delta G(V+\Delta V,B)\delta G(V,B)\rangle_B / 
\langle \delta G^2 \rangle_B ,$$ 
by means of averaging over different magnetic field points within the interval
0$<B<$1T. Then, the obtained correlation function is additionally averaged over
a narrow interval of bias voltage, not more than 2-3 times broader than the width
of the autocorrelation function determined after the first step. This procedure allows 
improved statistics and it slightly reduces variations in the form of the correlation function. 
Note that the finite amount of data used in this analysis still leaves space for statistical 
errors, so that the evaluated correlation function may be treated seriously only 
within an interval equal to 3 times its width at the half-maximum.

\begin{figure}
%
\epsfxsize=\hsize
\epsffile{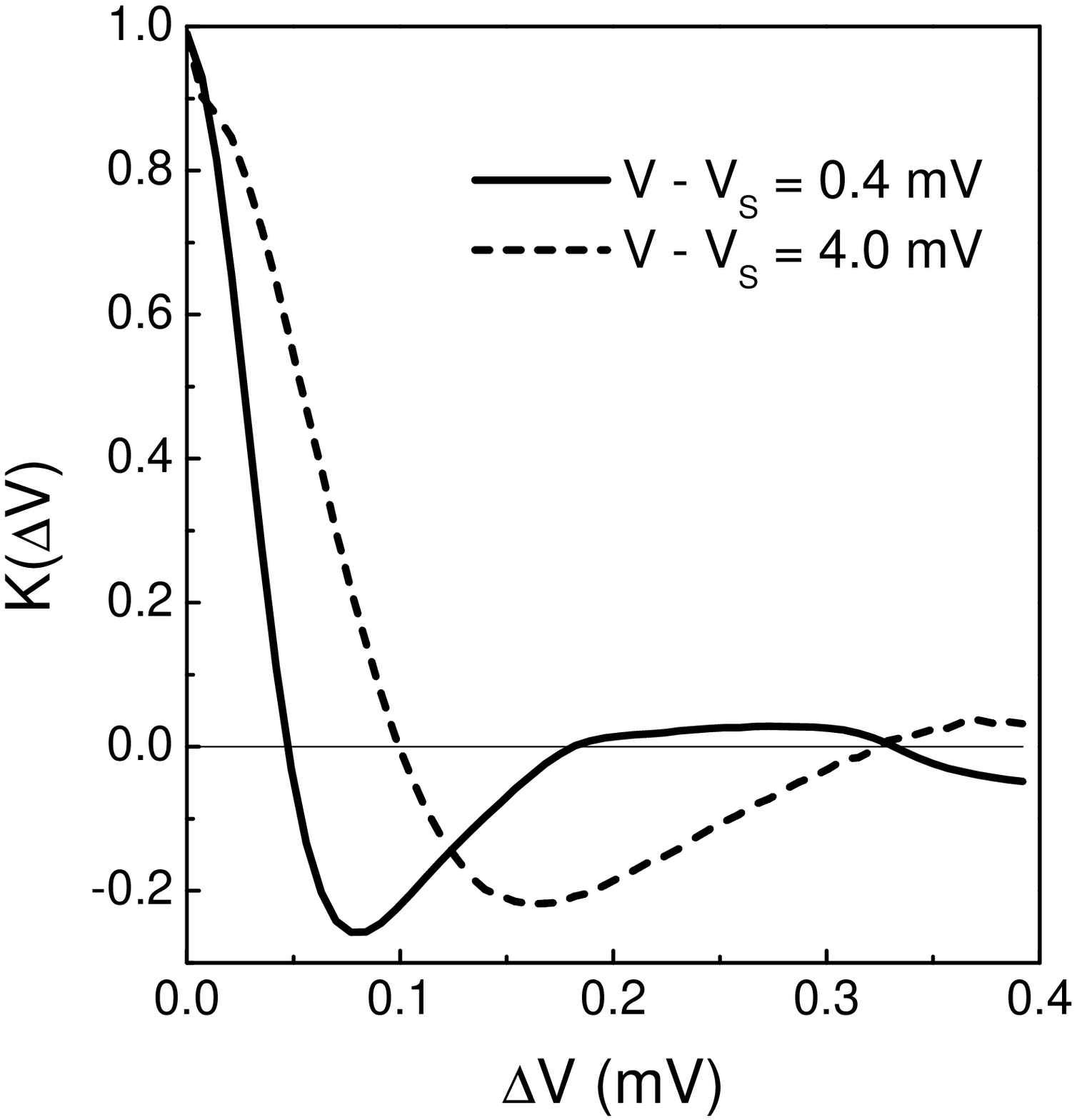}
\refstepcounter{figure}
\label{fig:corr_comp}
{\setlength{\baselineskip}{10pt} FIG.\ \ref{fig:corr_comp}.
Experimental correlation functions, taken at the beginning (solid, $V=10.2$~mV) 
and the end (dashed, $V=13.8$~mV) of the accessible voltage range.} 
\end{figure}

The typical result we get for such a correlation function is shown in 
Fig.~\ref{fig:corr_comp}
for two values of bias voltage: one at the beginning of the studied interval, at
$V=10.2$~mV, the other - at its end, at $V=13.8$~mV. These correlation functions have 
a very different width, which we attribute to an increase of inelastic broadening 
of states of quasi holes in the emitter upon the increase of their excitation energy, 
such that it becomes even larger than the spectrometer width, $\Gamma$. Therefore,
the comparison of correlation parameters of $K(\Delta V)$ can be used for determining 
directly the value of the inelastic relaxation rate of quasiparticles in the emitter as
a function of their excitation energy. 

To obtain an absolute value of the inelastic broadening from such a comparison,
one has to make a certain fit and, therefore, to use a certain form of the
correlation function $K(\Delta V)$. Theoretical analysis of the correlation function
of differential conductance fluctuations in Ref.~\onlinecite{fal97} has shown that its form, 
and, therefore, the value of the correlation voltage extracted from the fit depend on 
the effective dimensionality of the diffusive emitter, that is, on its geometry.  
In particular, for a quasi-0D emitter (diffusive pillar) and a quasi-2D film 
we have calculated
\begin{eqnarray}
K_0(\Delta V)= \frac{ 1-3\left(\frac{\Delta V}{V_c}\right)^2}
{\left[ 1+\left(\frac{\Delta V}{V_c}\right)^2 \right]^3}.
\label{cor0d} \\
K_2(\Delta V)= \frac{ 1-\left(\frac{\Delta V}{V_c}\right)^2}
{\left[ 1+\left(\frac{\Delta V}{V_c}\right)^2 \right]^2}.\nonumber
\end{eqnarray}
For a quasi-1D wire, and 3D bulk, these are, respectively,
\begin{eqnarray}\label{other-corr}
K_1(\Delta V)= \frac{(4-2Y-Y^2)\sqrt{1+Y}}{\sqrt{2}Y^5},\nonumber
\\
K_3(\Delta V)= \frac{(2-Y)\sqrt{1+Y})}{\sqrt{2}Y^3},
\; \nonumber \\ 
Y=\sqrt{1+(\Delta V/V_c)^2}.
\end{eqnarray}
All these correlation functions were obtained in the 
unitary symmetry class limit for fluctuations.

\begin{figure}
\epsfxsize=\hsize
\epsffile{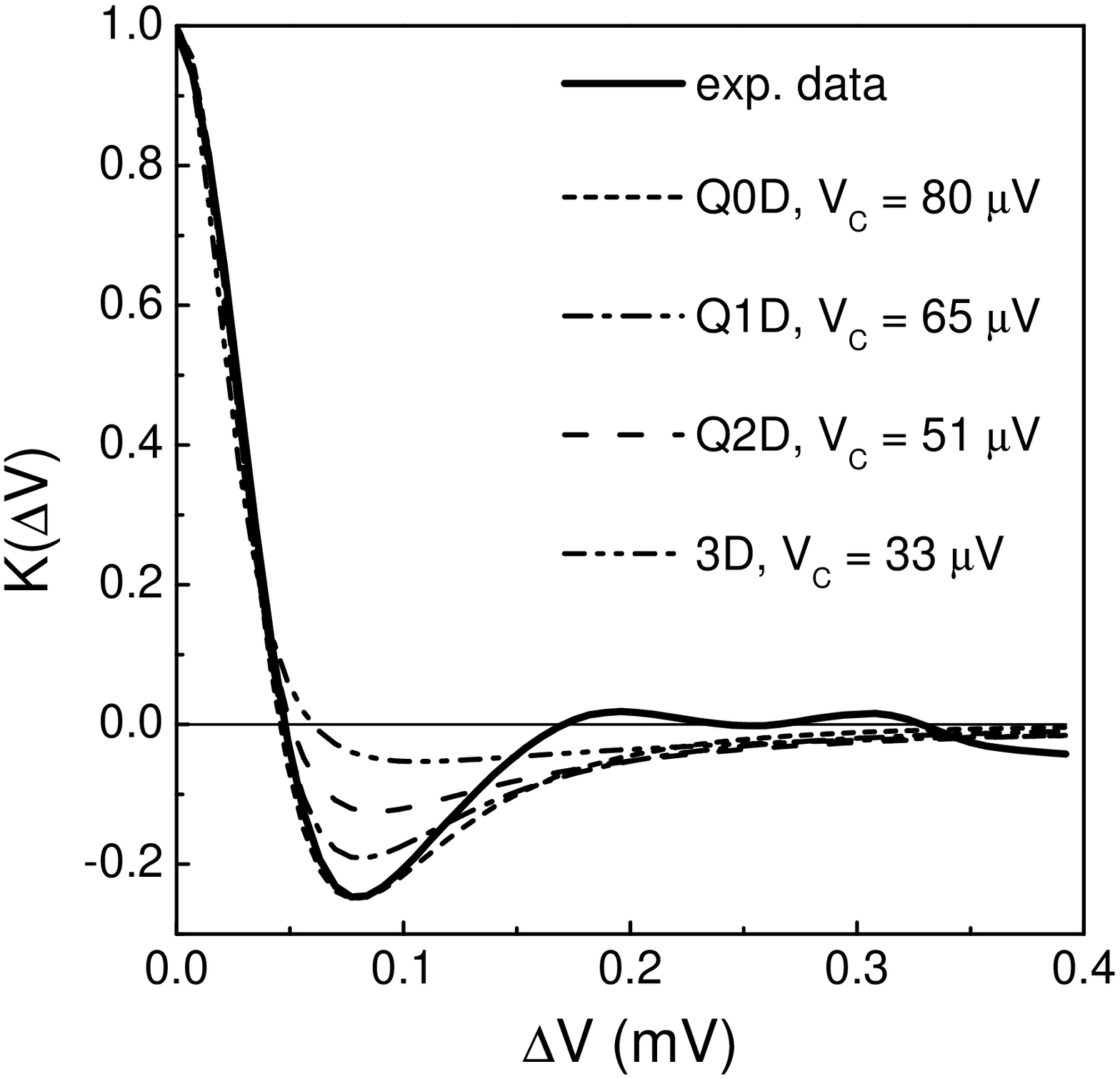}
\refstepcounter{figure}
\label{fig:corr_begin}
{\setlength{\baselineskip}{10pt} FIG.\ \ref{fig:corr_begin}.
The correlation function at the beginning of the accessible voltage range.
The solid line is the experimental correlation function at $V=9.8$~mV
and the other lines are fits based upon different assumptions about geometry.}
\end{figure}

In Fig.~\ref{fig:corr_begin}, all four 
of them are compared to the experimentally determined
correlation function for the smallest bias voltage interval, i.e., for $V=9.8$~mV. 
Theoretical curves shown in this plot for various models of an emitter can be 
characterized by the depth of a negative anti-correlation overshoot in $K$,
which is the most pronounced in the quasi-0D case.  
For each theoretical curve, the fit to the data is made using a single parameter, 
$V_c$, and the best agreement between the theory
and experimental data is achieved for the quasi-0D model of the emitter. 

The suggestion that the emitting electrode in the studied structure has the
form of a box, rather than the form of a wire which would be a natural assumption 
based upon the shape of the lithographically processed material in Fig.~\ref{fig:het}, 
needs an explanation. The point is that the emitter side of this device has been 
produced by overgrowing heavily doped GaAs:Si substrate ($10^{18}$~cm$^{-3}$ of Si) 
with a 300~nm buffer layer of GaAs:Si $4\times 10^{17}$~cm$^{-3}$. It is known
that the interface
between the substrate and the first grown layer is not as perfect as
the interfaces produced during the MBE growth process. It is 
expected that at the interface between the substrate and the first layer
a higher density of background impurities are incorporated and also that the
dislocation density will be higher than in the rest of the structure. 
The predominant background impurities will be carbon impurities, which
act as acceptors in GaAs and compensate the Si-donor doping.
Due to this compensation this interface could be poorly conducting. 
Although poor conduction through this interface does not affect the observable 
resistance value of the device and no other measurement 
performed on structures from the same series had enough sensitivity to
indicate its presence, the LDOS fluctuations measurements appear to be sensitive
enough to illuminate its existence. 

For each given geometrical shape of the emitter (for this sample, a  
$L$-thick disk with a radius $R$), the 
effective dimensionality reflected by the shape of the correlation function in
Eq.~(\ref{cor0d}) also depends on the ratio between the diffusion length, 
\begin{equation}
\label{Lc}
L_c =\sqrt{\frac{\hbar D}{\Gamma  +\hbar \gamma}}
\end{equation}
and geometrical sizes, $L$ and $R$. The diffusion length in Eq.~(\ref{Lc}) characterizes
the volume of a disordered system that is effectively tested by a coherently diffusing particle 
within the time scale taken before it either escapes from the emitting electrode 
to the collector via the resonant impurity S or relaxes inelastically into states
at different energies. When the latter length scale is the largest,  $L_c\gg L,R$, 
the correlation function of fluctuations has the quasi-0D form. When $R>L_c >L$, the finite 
radius of a pillar would not matter, and the correlation function would have the quasi-2D 
form. Similarly, $L>L_c >R$ would correspond to the quasi-1D result in Eq.~(\ref{cor0d}). 
Finally, one would have to treat the regime of $R,L\gg L_c$ as the three-dimensional one. 

The value of the correlation voltage extracted from the fit of the experimental data 
in Fig.~\ref{fig:corr_begin} using the quasi-0D model, $V_c=80$~$\mu$V is very close 
to the
width of the main resonance peak in Fig.~\ref{fig:gv_av} determined by the intrinsic 
spectrometer width, $V_\Gamma =72$~$\mu$V. Comparison of $V_\Gamma$ with other values of 
$V_c$ obtained from fits of experimental $K(V)$ 
using other dimensionality assumptions [$V_c=$ $65$~$\mu$V, $51$~$\mu$V and $33$~$\mu$V for the Q1D, Q2D 
and 3D models, respectively] gives an additional argument in favor of the view that 
we deal here with a quasi-0D emitter. At the same time, the relevant diffusion length 
$L_c$ calculated as $L_c =\sqrt{D\hbar/E_c} \approx \sqrt{D\hbar /\Gamma}$ 
is longer than both
the sample diameter and the width of the buffer layer, which would be consistent with
an assumption that the interface is an obstacle for electron escape to the substrate.

\begin{figure}
\epsfxsize=0.6\hsize
\epsffile{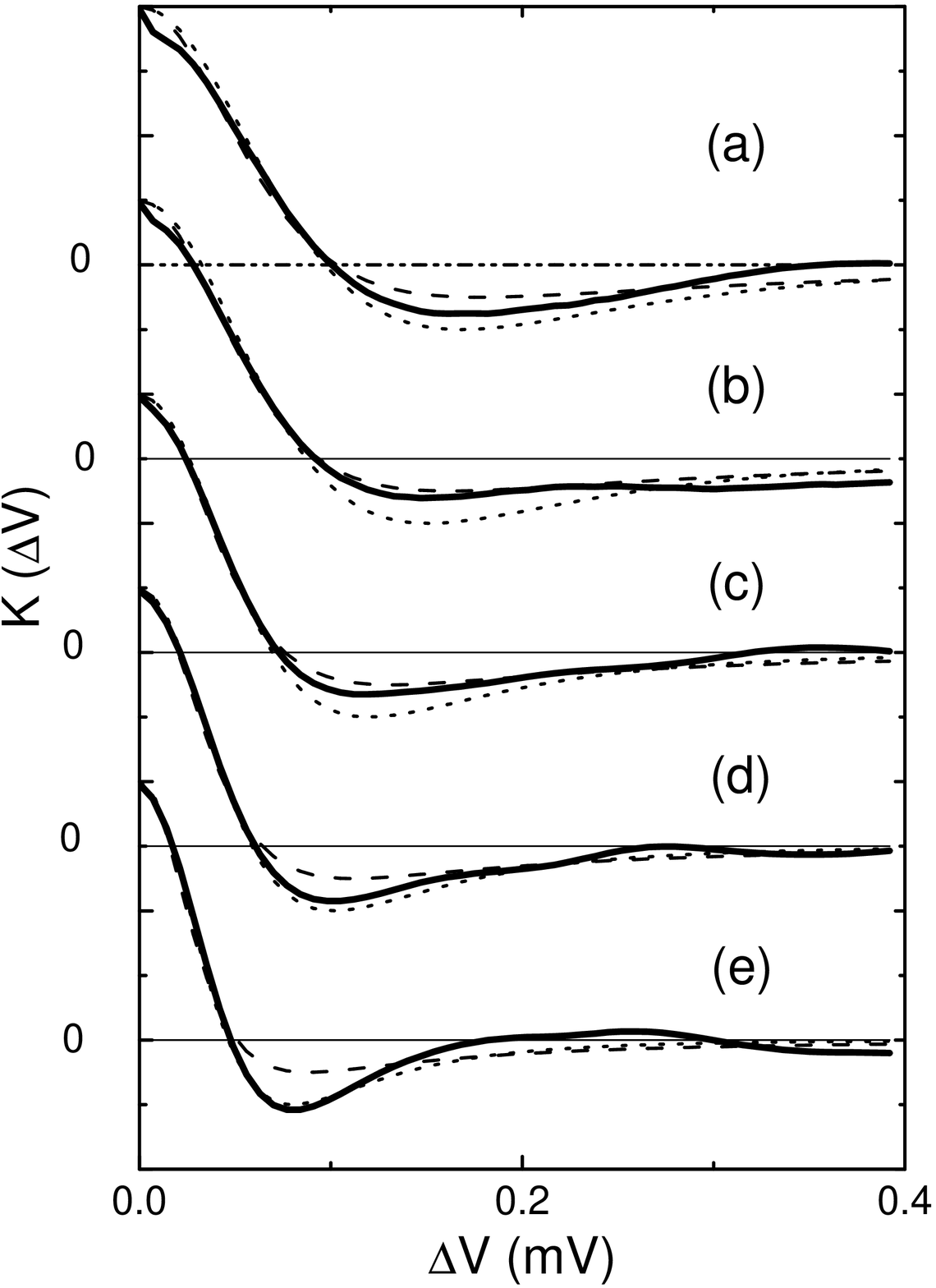}
\refstepcounter{figure}
\label{fig:corr_table}
{\setlength{\baselineskip}{10pt} FIG.\ \ref{fig:corr_table}.
The evolution of the correlation function from the beginning (bottom)
to the end (top) of the accessible voltage range
showing the experimental correlation function (solid) 
and fits for the quasi-0D (dotted) and quasi-2D model (dashed) for five voltages: 
(a)$V=13.8$~mV, (b)$V=13.0$~mV, (c)$V=12.2$~mV, (d)$V=11.4$~mV and (e)$V=10.6$~mV.} 
\end{figure}

Since the length scale $L_c$ in Eq.~(\ref{Lc}) shortens, due to faster inelastic
relaxation as the quasi-particle excitation energy increases, the effective dimensionality 
of the system may vary across the bias voltage interval we study. Since the sample used 
here has $R>L$, a crossover may take place between the quasi-0D and quasi-2D form 
of the correlation function that should be used for fitting the data in the broader 
voltage interval. One can find indications of such a crossover in the series 
of correlation functions shown in Fig.~\ref{fig:corr_table}. 



Traces of crossover behavior 
in Fig.~\ref{fig:corr_table} require one to make a detailed 
theoretical analysis of the intermediate regime
$L_c \approx R$, since our final goal is to obtain quantitative information 
about the quasi-particle lifetime, as a function of quasiparticle energy in the
entire energy interval assessed in the reported measurement. Details
of a calculation of correlation functions in the crossover regime are presented 
in Appendix B. Here, we only describe the results, in a graphic form.  
Fig.~\ref{fig:thfig5} shows the change in the shape of the correlation function
of differential conductance fluctuations expected for a spectrometer 
placed in the center of the bottom surface of a round disk, for various values 
of the ratio $L_c /R$, but for the same nominal $V_c$. This plot shows that
the crossover between the quasi-0D (dotted bottom line) and quasi-2D form 
(solid line) can be split into two steps. 
First, the negative deep in $K(V)$ at $V \approx V_c$ is reduced 
(anti-correlations become weaker), which happens without a noticeable
change in the width of the correlation function at the half maximum
(in units of $V/V_c$). 
The following evolution of the form consists of the broadening 
of the main part of the correlation function. This two-step evolution 
suggests that the fit to the central peak of the experimentally determined 
correlation function using the quasi-0D formula is a consistent procedure 
applicable even across some part of the crossover regime. The need for such a  
simplified procedure in the following analysis has another reason.
When the crossover takes place, the exact form of $K(V)$ becomes dependent on 
the position of the spectrometer on the surface, that is, its distance to the 
disk perimeter. This effect is illustrated in the inset to Fig.~\ref{fig:thfig5}
using several plots of $K(\Delta V)$ calculated for different off-center 
positions of the resonant impurity. Plots in Fig.~\ref{fig:thfig5}  
also show that values of the sum of the physical parameters, 
$\Gamma +\hbar\gamma$ obtained following such a procedure may be 
overestimated when the crossover to the quasi-2D limit is 
more developed.

\subsection{Analysis of the variance of differential conductance fluctuations}

\begin{figure}
%
\epsfxsize=\hsize
\epsffile{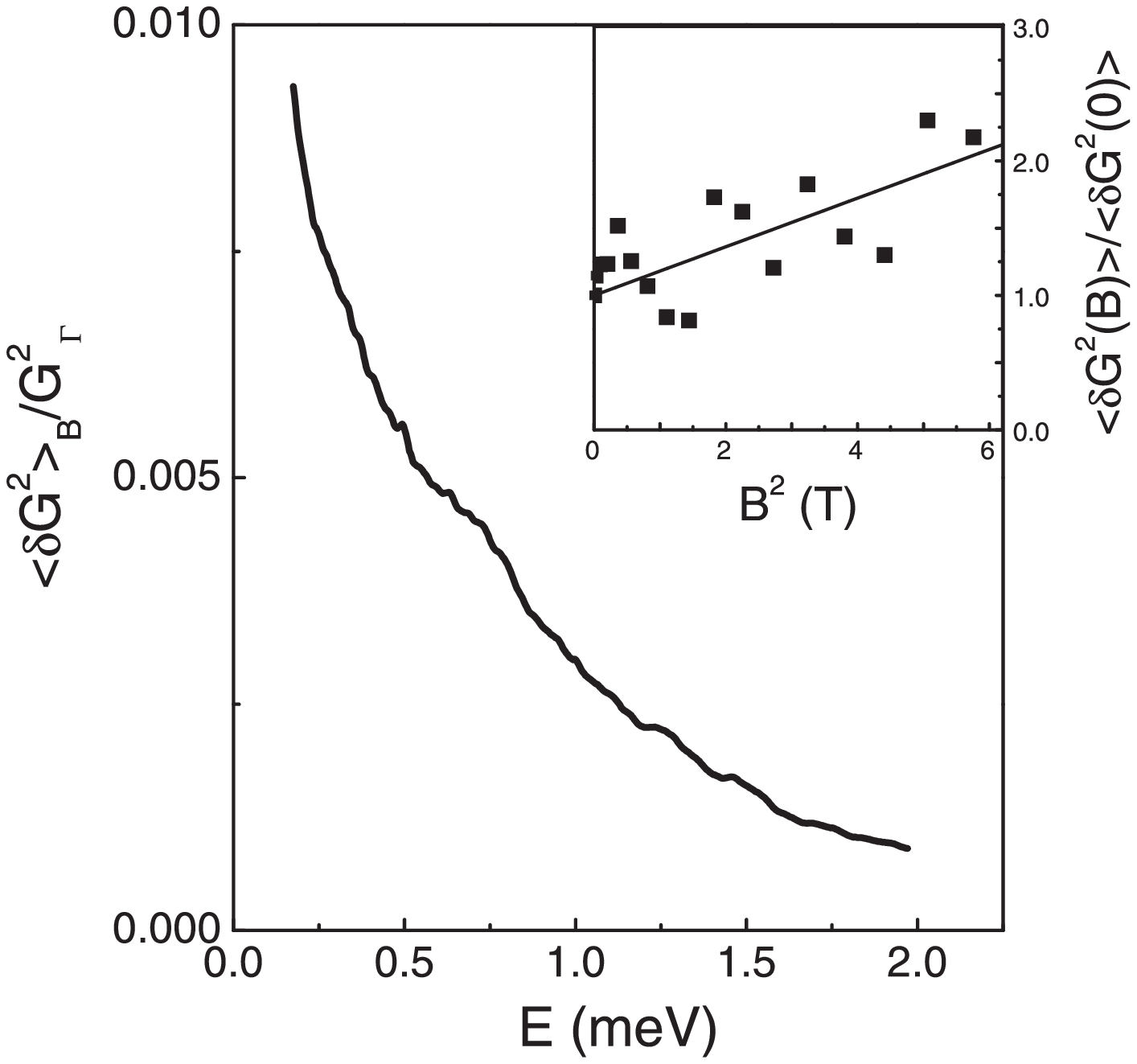}
\refstepcounter{figure}
\label{fig:fluct_rate}
{\setlength{\baselineskip}{10pt} FIG.\ \ref{fig:fluct_rate}.
Inelastic quasi-particle relaxation $\gamma(E)$: Variance of the differential 
 conductance $\delta G(E)^2$ versus excitation energy $E$.
The inset shows the increased variance of fluctuations in classically
high magnetic fields $\omega_c \tau \sim 1$.}
\end{figure}

Quantitative information about the energy dependence of 
inelastic quasi-particle relaxation 
can also be extracted from the bias voltage dependence of the variance
of differential conductance fluctuations. Such a dependence for 
the sample described in this paper is
shown in Fig.~\ref{fig:fluct_rate}. It is evaluated on the basis of the pattern of 
raw data in Fig.\ref{fig:map} after subtracting from the data the
average conductance, $G(V)$, shown in Fig.\ref{fig:gv_av}, then, 
averaging the difference over the  magnetic field interval 0$<B<$1T 
\begin{equation}\label{varb}
{\rm var}_BG = 
\int_{0{\rm T}}^{1{\rm T}}\frac{dB}{1{\rm T}}
\left( G(V,B)-\langle G(V)\rangle_B \right)^2,
\end{equation}
and, then, by 
smoothing it over the bias voltage interval of 3 times $V_c$
determined for the corresponding bias voltage range in the previous section.
The result is presented in the form normalized by the height of the
main conductance peak, $G_{\Gamma}$, in order to exclude from this
analysis the parameters of tunneling barriers, and the bias voltage value
is converted here into the excitation energy of a quasi-particle ($E$ is the
energy of the Fermi sea hole evaluated with respect to the Fermi level). 
Because of the above mentioned smoothing procedure, we cannot start the plot in 
Fig.~\ref{fig:fluct_rate} from exactly $E=0$.

The decrease of the amplitude of differential conductance fluctuations 
upon the increase of excitation energy of quasi-particles is 
attributed to a faster inelastic relaxation of the latter, which can 
be used to study the dependence $\gamma (E)$. Similarly to the correlation
function, the exact form of such a dependence varies if one makes 
different assumptions about the effective dimensionality, $d$: 
\begin{equation} \label{variance}
\frac{\langle \delta G^2 \rangle }{G_\Gamma^2} =
\frac{1}{\left[1+\frac{\hbar\gamma}{\Gamma}\right]^{3-d/2}} 
\times \left\{
\begin{array}{r@{\quad}l}
\frac{1/2}{\nu L R^2\Gamma}, & Q0D \\
& \\
\frac{3/16}{\nu R^2\sqrt{\hbar D\Gamma}}, & Q1D \\
&  \\
\frac{1/16}{\nu \hbar DL}, & Q2D \\
& \\
\frac{\sqrt{\Gamma /\hbar D}}{32 \nu \hbar D}, & 3D. 
\end{array}
\right.
\end{equation}

Using the measured spectrometer width $\Gamma$ and the known sample
dimensions $R$ and $L$, these equations enable us to obtain
theoretical estimates of the amplitude of the variance for a
given effective dimensionality.
A comparison with the low $E$ part of
the measured variance data plotted in Fig.~\ref{fig:fluct_rate},
where we expect $\gamma (E) \sim 0$, has the best agreement with
the Q0D theory.
In these estimations we used the value of the mean free path,
$l\approx 70$~nm, assigned to the nominal
doping level of the buffer layer.
This value of $l$ is confirmed by the tendency of the variance
$\langle \delta G(B)^2\rangle$ to follow a
$[1+(\omega_c \tau )^2]$-dependence\cite{fal97,hol00} 
at classically high magnetic fields $\omega_c \tau \sim 1$
as shown in the inset of Fig.~\ref{fig:fluct_rate}.

\begin{figure}
\epsfxsize=\hsize
\epsffile{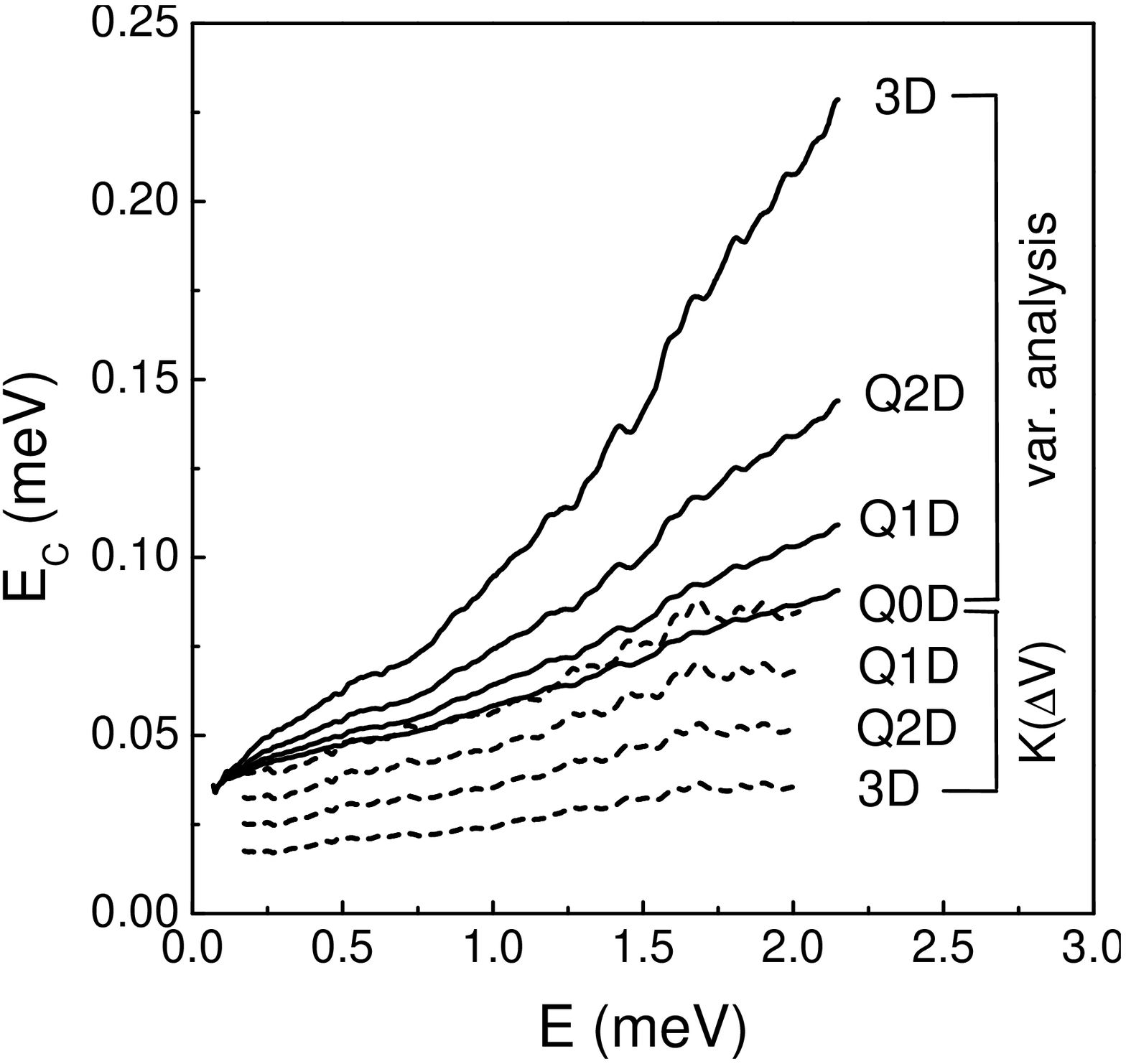}
\refstepcounter{figure}
\label{fig:fluctcorr_fit}
{\setlength{\baselineskip}{10pt} FIG.\ \ref{fig:fluctcorr_fit}.
Comparison of the correlation energy $E_c$ of LDOS fluctuations
extracted from the amplitude [solid lines] and the correlation
function $K (\Delta V)$ [dashed lines] for
different models of quasi-dimensionality.} 
\end{figure}

The energy dependence of the parameter $E_c =\Gamma +\hbar\gamma$
can be extracted from Fig.~\ref{fig:fluct_rate} using the
formulae in Eq.~(\ref{variance}) and it is plotted in
Fig.~\ref{fig:fluctcorr_fit} for the four different effective
dimensionalities (upper, solid lines).
Also plotted in Fig.~\ref{fig:fluctcorr_fit} is the energy
dependence of the parameter $E_c =\alpha V_c$ obtained from the
analysis of the correlation function (lower, dashed lines).
The values of $E_c$ obtained along two different roots have to 
coincide for an appropriate dimensionality assumption, and they 
agree only when analysis is based upon the quasi-0D emitter model.




\subsection{Quasi-particle inelastic relaxation rate in a disordered conductor.}
\begin{figure}
%
\vspace*{0.25cm}
\epsfxsize=\hsize
\epsffile{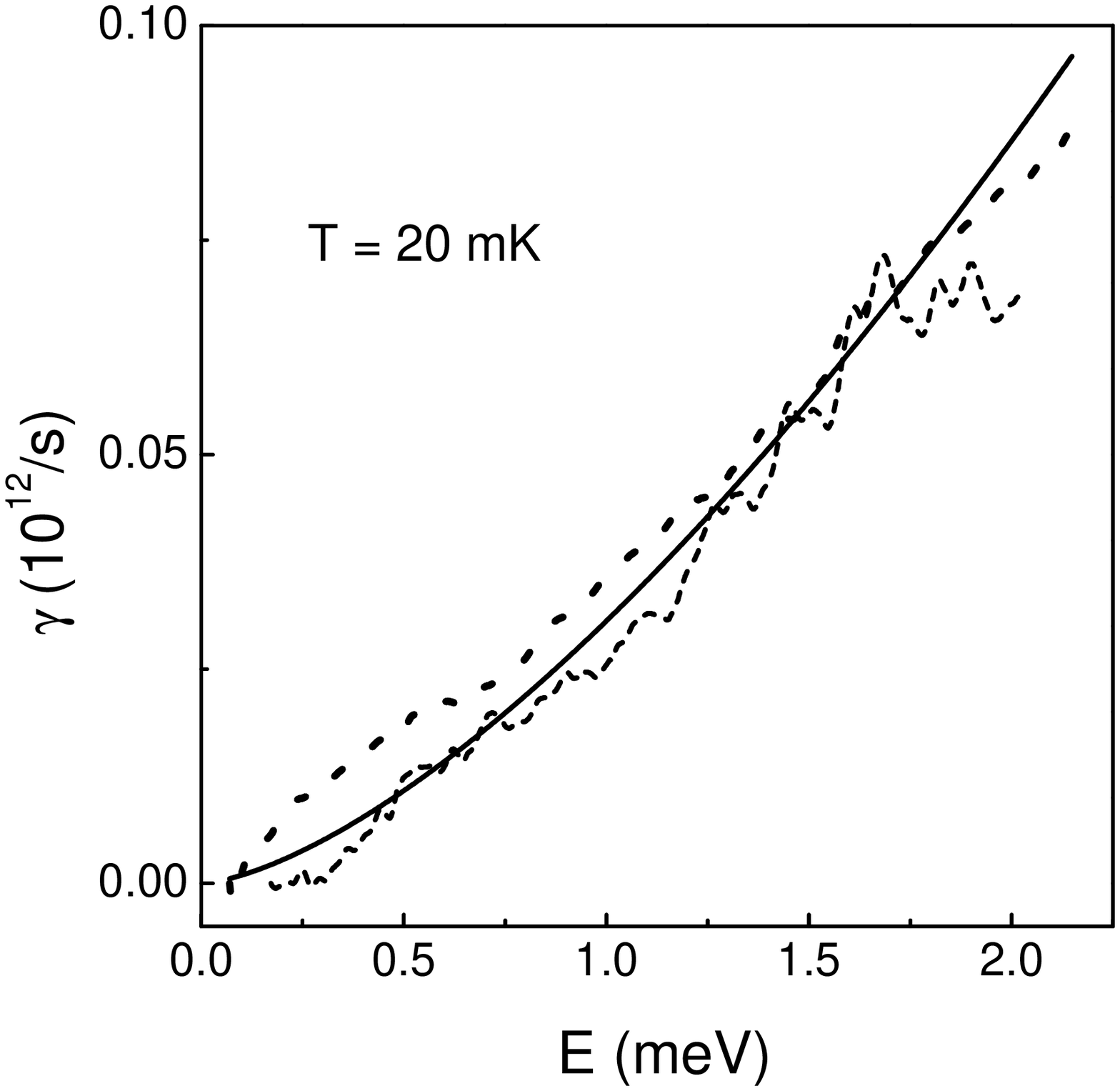}
%
\refstepcounter{figure}
\label{fig:corrfluct_fit0d}
\vspace*{0.25cm}
{\setlength{\baselineskip}{10pt} FIG.\ \ref{fig:corrfluct_fit0d}.
Determination of the quasiparticle relaxation rate from analysis of
correlation (dotted line) and fluctuation data (dashed line).
The solid line is a fit to the theoretically expected inelastic particle
relaxation rate $\gamma(E)$, see text for details.}
\end{figure}

%
On the basis of the  material presented above, we conclude that the 
use of the quasi-0D assumption for the analysis of fluctuations is fully justified
and can be exploited for analyzing the energy dependence of the inelastic 
relaxation rate of quasi-particles, $\gamma (E)$. The latter can be obtained from 
the data shown in Fig.~\ref{fig:fluctcorr_fit} by subtracting the original 
spectrometer width. 
The resulting relaxation rate dependence on the excitation energy is shown
in Fig.~\ref{fig:corrfluct_fit0d}.  This plot contains two sets of data taken from the 
analysis of correlation functions and the variance, and the comparison to the 
rate values calculated using Altshuler-Aronov theory.  The discrepancy between 
data worked out in two different ways indicate the arrow bars one would have to 
assign to the presented analysis.


%



The theoretical curve shown in Fig.~\ref{fig:corrfluct_fit0d} is a fit
to the relaxation rate as derived by Sivan, Imry and Aronov,\cite{siv94} 
using 
$E_F=30$~meV for the emitter buffer doped to  $4.0\times 10^{17}$~cm$^{-3}$
with Si,
\begin{equation}\label{relaxationrate}
\gamma = \frac{105\sqrt{3}}{16\pi} \frac{\hbar^{1/2}E^{3/2}}{\tau^{3/2}E_F^2}.
\end{equation}
The mean free path obtained from this fit is $l = 93$~nm, which is close to
mean free path expected for this nominal doping (between $l = 50$~nm and
$l = 100$~nm) and also close to the value extracted from the analysis of
the increase of the variance of fluctuations with magnetic field ($l = 70$~nm).
The use of the three-dimensional expression for the relaxation rate in 
Eq.~(\ref{relaxationrate}), in contrast to the quasi-0D model used
to describe fluctuations, is justified by the following reason. As discussed at the
end of Appendix A, the relaxation of a quasi-particle with energy $E$ 
is dominated by  
electron-electron (e-e) collisions with energy transfer comparable 
to $E$, and such a rate is determined by correlations between
chaotic wavefunctions with a typical energy separation $\epsilon\approx E$.
The latter are formed at the length scale $L_\epsilon\approx\sqrt{\hbar D/\epsilon}$, 
which has to be compared to the system size: the pillar radius, $R$, and the 
width, $L$.  In particular, if $L_\epsilon <L<R$, the e-e interaction can be treated as
in the three-dimensional bulk of a disordered conductor. This condition 
can be expressed more rigorously as $E>\pi^2\hbar D/L^2, \pi^2\hbar D/R^2$
which states that the quasi-particle excitation energy has to be larger than 
the Thouless energy related to diffusive motion 
across the pillar. Since the extracted values $\gamma (E)$ sufficiently 
exceed experimental uncertainty only for quasi-particle excitation energies 
$E>0.5$~meV [which has to be compared to $\pi^2\hbar D/L^2 \approx 0.4$~meV], 
their quantitative comparison to the calculation of $\gamma (E)$ in the 3D limit 
seems to be consistent. At the same time, the entire interval 
of energies analyzed in Fig.~\ref{fig:corrfluct_fit0d} belong to a
clearly diffusive regime, $E<\hbar\tau \approx 4$~meV. 
Note that all this analysis is extended only over the 
low magnetic field range, where the Landau quantization of emitter
states does not play any role.

\section{Summary}
We study resonant tunneling through a discrete localized level in a GaAs/AlGaAs
double-barrier heterostructure. The differential conductance exhibits a 
temperature-insensitive fine structure which is attributed to 
fluctuations in the local density of states in the doped GaAs emitter.
The observed fine structure is analyzed in terms of variance of
the fluctuations in the differential conductance and in terms
of correlation functions with respect to voltage. From analyzing
the shape of the correlation function we conclude that the effective
dimensionality of the emitter is zero-dimensional caused by
the disordered interface between the GaAs substrate and the doped
buffer layer. In this experiment the electrons tunnel from below
the Fermi energy in the heavily doped emitter contact through
the discrete localized level leaving behind a quasi-hole  in the 
emitter. By quantitatively analyzing the width of the measured correlation
functions and the measured variance we are able to extract the energy dependence
of the inelastic quasi-hole relaxation. 
%

%
%
\begin{acknowledgments}
We thank A. F\"orster and H. L\"uth for growing the double-barrier heterostructure.
We acknowledge financial support from
BMBF, DFG, EPSRC, NATO, and TMR.
\end{acknowledgments}


\section*{Appendix A}

This appendix presents a qualitative method of estimating
the variance of differential conductance fluctuations
and the energy dependence of quasi-particle  relaxation.
It is constructed using a scaling picture similar to that of
Thouless\cite{tho77} by considering what happens to the states of
single electrons in a box when the electrons are able to diffuse
into other, similar boxes.
For clarity we begin by considering three dimensions $d=3$ although
this is not necessary for the following arguments to hold.
In a classical picture of diffusion, a diffusive path can be viewed
as a series of straight line segments of typical length
equal to the elastic mean free path
$l$, where $l = v_F\tau$, $\tau$ is the elastic time,
and $v_F$ is the Fermi velocity.
The classical diffusion coefficient is $D = v_F^2\tau/d$
and the typical time required to diffuse a length $\xi$
is $\tau_D (\xi )= \xi^2/D$.

Consider eight cubes of length $\xi$ which are separated
by barriers such that no particles may move between the
cubes.
We imagine that it is possible to diagonalize the
Hamiltonians of the
separate cubes
and we denote the eigenstates as
$\psi_{\alpha i}^{\xi} (r)$
where $\alpha$ specifies the cube
and $i$ specifies the state.
These states are called `mother states' of the
generation $\xi$ and they have a mean level
spacing $\Delta (\xi )$ where $\Delta (\xi ) = 1/(\nu \xi^d)$.
The states are normalized so that
$\int |\psi_{\alpha i}^{\xi} (r)|^2 dr = 1$.
The Hamiltonian of the total system, consisting of eight
cubes of size $\xi$ with barriers between them, is also
diagonal.

When the barriers between the cubes are removed,
particles may diffuse between them.
As stated above, the typical time to diffuse a length $\xi$
is $\tau_D (\xi )= \xi^2/D$.
The energy corresponding to this time is called
the Thouless energy
$E(\xi ) \approx h D/\xi^2$.
Diffusion between cubes produces a finite
mixing of states from different Hamiltonians so that
the Hamiltonian of the new system (consisting of
the eight smaller cubes) is not diagonal.
Instead it has finite elements within a distance
$E(\xi )$ of the main diagonal and has elements
which are approximately equal to zero elsewhere.

An approximation is used to diagonalize the new Hamiltonian.
An area of width $E(\xi )$ is centered
on the middle of the Hamiltonian and
a unitary transformation $U$ is applied to diagonalize it,
neglecting the rest of the Hamiltonian.
Approximate eigenstates of the new system are linear
combinations of a finite number of mother states
of the generation $\xi$,
\begin{equation}
\psi_{\beta n}^{2 \xi} ({\bf r})
\approx
\sum_{\alpha i} a_{\alpha i}^{\beta n} \psi_{\alpha i}^{\xi} ({\bf r})
\label{lincom}
\end{equation}
where $a_{\alpha i}^{\beta n}$ are coefficients with indices
$\alpha i$ that refer to the original cubes of scale $\xi$
and index $n$ of the new states in the cube $\beta$
of scale $2\xi$.
The new approximate eigenstates are normalized
so that
\begin{equation}
\sum_{n} |a_{\alpha i}^{\beta n}|^2 \equiv
\sum_{\alpha i} |a_{\alpha i}^{\beta n}|^2 = 1
\label{coeffnorm}
\end{equation}
where we used the property $U^{\dagger}U =1$
(only one value of $\beta$ is considered).
Correlations between local densities remain important because
at each level in the scaling procedure there is only a finite basis
involved in the construction of new states.

Consider an experimental observation of the local density of
states at position ${\bf r_0}$.
Formally the local density of states may be expressed
in terms of a summation of states.
When observing through a spectrometer of energy width
$\Gamma$ then this may be written as a sum of states with
energy $E_i$ within $\Gamma$ of the spectrometer energy $E_S$,
\begin{equation}
\nu ({\bf r_0} ; E_S )
\approx
\Gamma^{-1}
\sum_{|E_i - E_S|\leq \Gamma}
\left| \psi_{\alpha i}^{\xi} ({\bf r_0}) \right|^2 ;
\quad \xi \leq L_{\Gamma} .
\label{ldos1}
\end{equation}
The approximate eigenstates
$\psi_{\alpha i}^{\xi} ({\bf r_0})$ are those in a
system of size $\xi$ where $\xi \leq L_{\Gamma}$ and
$L_{\Gamma}$ is the
length scale corresponding to energy $\Gamma$,
$L_{\Gamma} \approx \sqrt{h D/\Gamma}$.
In general, however, the system is larger than $L_{\Gamma}$
and it is necessary to know how the above summation
behaves at larger scales $\xi > L_{\Gamma}$.
Consider for example a system of scale $\xi = 2L_{\Gamma}$.
Applying the above scaling procedure we may write the
approximate eigenstates at larger scales using Eq.(\ref{lincom}),
giving
\begin{eqnarray}
\nu ({\bf r_0} ; E_S ) &\approx&
\Gamma^{-1}
\sum_{|E_i - E_S|\leq \Gamma}
\left| \psi_{\alpha i}^{\xi =2L_{\Gamma}} ({\bf r_0}) \right|^2
\nonumber \\
&\approx&
\Gamma^{-1}
\sum_{n} \sum_{\alpha i}
|a_{\alpha i}^{\beta n}|^2
\left| \psi_{\alpha i}^{L_{\Gamma}} ({\bf r_0}) \right|^2
\nonumber \\
&\approx&
\Gamma^{-1}
\sum_{\alpha i}
\left| \psi_{\alpha i}^{L_{\Gamma}} ({\bf r_0}) \right|^2
\label{ldos2}
\end{eqnarray}
where in the last step we used the normalization condition
given in Eq.(\ref{coeffnorm}).
This result shows that the summation over the energy interval
$\Gamma$ will not vary when the spectrum is modified
into its final form at the total system size $L$, but it will
depend on the spectrum at length scale $L_{\Gamma}$.
This is because once the scale $\xi > L_{\Gamma}$ then the
corresponding energy $E(\xi ) < \Gamma$ and information
about correlations that is carried by the mother states will
remain in the energy interval $\Gamma$ no matter how large
$\xi$ becomes.
Now we describe the application of the above scaling picture
to the differential conductance.\cite{fal97,hol00} 
The current in the plateau regime, $I$, is determined by a sum of local
densities of the wavefunctions
$\left| \psi_{E} ({\bf r}) \right|^2$ 
with energy, $E$, taken in an energy interval $\Gamma$
around the energy $E_0$,
$I \propto \nu \sim
\Gamma^{-1} \sum \left| \psi_{E} ({\bf r}) \right|^2$.
The number of states in a sample of volume $L^d$ within the energy
interval $\Gamma$  is $N(\Gamma ,L) \approx \nu_0 \Gamma L^d$ where
$\nu_0$ is the mean density of states per unit volume, per unit energy.
The variance of the differential conductance, 
$\left< \delta G^2\right>$, is given by a typical
fluctuation in the density of states $\delta \nu$ divided by the
typical energy interval $\Gamma$. 
As described above, a summation over the energy interval
$\Gamma$ depends on the spectrum at length scale $L_{\Gamma}$
with number of states $N (\Gamma ,L_{\Gamma})$.\cite{fal97,hol00} 
Since $\psi_{E} ({\bf r})$ from
a single state is a random variable with mainly Gaussian statistics
in the metallic regime\cite{F+E96} and the variance,
$\left< \delta G^2\right>$, is given by a
sum of the individual variances, we have
$$\left< \delta G^2\right>
\sim
\frac{N(\Gamma ,L_{\Gamma})
\left< \left| \psi_{E} ({\bf r}) \right|^2 \right>^2}
{\Gamma^{2} V_{\Gamma}^2} , $$
where the typical density of a single state is
$\left< \left| \psi_{E} ({\bf r}) \right|^2 \right>
\sim 1/L_{\Gamma}^d$.

In the above estimation $V_{\Gamma}$ is the smallest voltage step
which is given by the spectrometer width $V_{\Gamma} \sim \Gamma /e$.
An additional level broadening, $\hbar\gamma$, takes into account
relaxation processes in the bulk of the emitter and
results in a total level broadening
$\Gamma \left(1 + \hbar\gamma /\Gamma \right)$.
Counting powers of $\Gamma$ in the estimation of
$\left< \delta G^2\right>$ at the end of the last paragraph leads to
a factor of $\left(1 + \hbar\gamma /\Gamma \right)^{d/2-3}$
in the variance.
However we should stress that the level broadening $\hbar\gamma$
does not influence the average value of the current in the plateau regime,
$$\left< I \right>
\sim
\frac{N(\Gamma ,L_{\Gamma}) 
\left< \left| \psi_{E} ({\bf r}) \right|^2 \right>}
{\Gamma},$$
or the height of the main differential peak which is given by the mean current
divided by the width of the peak,
$$G_{\Gamma}
\sim
\frac{N(\Gamma ,L_{\Gamma}) 
\left< \left| \psi_{E} ({\bf r}) \right|^2 \right>}
{\Gamma V_{\Gamma}}.$$
Thus the variance is parametrically reduced as compared to $G_{\Gamma}^2$
by a factor $1/N(\Gamma ,L_{\Gamma})$ 
where
$N(\Gamma ,L_{\Gamma})
\sim
g (L_{\Gamma})
\sim
\nu D^{d/2} \Gamma^{1-d/2}.$
When the dimension $d$ refers to the effective dimensionality
of the system as determined by the volume over which mesoscopic
fluctuations occur, embedded in a nominally three dimensional space,
then the factor $N(\Gamma ,L_{\Gamma})$ is replaced by
$N(\Gamma ,L_{\Gamma}) L^{3-d}$.
In this case the variance may be written as
$$\left< \delta G^2\right>
\sim
\frac{1}{\nu D^{d/2} \Gamma^{1-d/2} L^{3-d}}
\frac{G_{\Gamma}^2}
{\left(1 + \hbar\gamma /\Gamma \right)^{3 - d/2}}.
$$

It is also possible to explain the energy dependence
of the inelastic scattering rate
in terms of the above scaling picture.
The rate is determined by a collision between four particles
involving transferred energy $\omega$.
There are two initial particles
with energies $E > 0$ and $\epsilon^{\prime} < 0$ and two final
particles with energies $E - \omega > 0$ and
$\epsilon^{\prime} + \omega > 0$.
The inelastic rate may be estimated using Fermi's Golden Rule\cite{siv94,bla96,alt97,alt98}
\begin{eqnarray}
\gamma (\xi )
\sim \sum_{0<\omega <E} \, \,\sum_{-\omega<\epsilon^{\prime} <0}
\frac{|M (E ,\epsilon^{\prime}, \omega )|^2}{\Delta (\xi )},
\nonumber
\end{eqnarray}
where $M (E ,\epsilon^{\prime}, \omega )$ is the matrix
element for the collision.
For a short ranged interaction the matrix elements are given by
a spatial integration of a product of four single particle
wave functions\cite{bla96,alt97}
\begin{eqnarray}
\frac{M (E ,\epsilon^{\prime}, \omega )}{\Delta (\xi )}
\approx
\xi^d \int d^dr \, \, 
\psi_{\epsilon^{\prime} + \omega}^{*} ({\bf r})
\psi_{E - \omega}^{*} ({\bf r})
\psi_{\epsilon^{\prime}} ({\bf r})\psi_{E} ({\bf r}) .
\nonumber
\end{eqnarray}
The normalized wavefunctions in a disordered system exhibit random spatial
oscillations, each typically contributing
$|\psi |^2 \sim \xi^{-d}$.
After disorder and spatial averaging the product of four wavefunctions is
roughly
$$\left<\psi^4 \right>
\sim
\frac{1}{\xi^{2d} N (E(\xi ),\xi)}.$$
In the limit $N (E(\xi ),\xi) \rightarrow \infty$,
there is an infinite basis involved
in the construction of new states at each level of the scaling process,
leading to an absence of correlations between different eigenvectors.
For finite $N (E(\xi ),\xi)$, however,
there is a finite basis and correlations exist.
Integration over the hypercube provides an additional factor of $\xi^d$
so that a typical value of the matrix element is
$$M
\sim
\frac{\Delta (\xi )}{N (E(\xi ),\xi)}.$$
Since each summation with respect to energy contributes
roughly $E /\Delta (\xi )$
we find
$$\gamma (\xi )
\sim
\frac{E^2}{N^2 (E(\xi ),\xi) \, \Delta (\xi )} .$$

Now we consider what occurs when the hypercubes are scaled up to the
total system size $L$.
For small energies, $E < E(L) \equiv E_{Th}$, the system is in the
zero dimensional limit whereby the system size is always less than
the length scale associated with the energy, $L_{E}$, where
$L_{E} \approx \sqrt{h D /E}$.
The above estimation holds for all scales up to the system size 
so that the inelastic rate is\cite{siv94,alt98}
$\gamma \sim E^2 /(g^2(L)\Delta (L))$
where $g(L) \sim N (E_{Th},L)$.
For our case of interest, however, we sum over states up to
energies greater than the Thouless energy $E_{Th}$ of the total system,
$E > E(L) \equiv E_{Th}$.
When the system size reaches $L_{E}$, the energy scale
$E$ contains all the information about correlations between
the states.
As for the calculation of $\left< \delta G^2\right>$
at the length scale $L_{\Gamma}$, further scaling does not change
the evaluation of the relaxation rate.
The summation over the energy interval
$E$ will not vary when the spectrum is modified
into its final form at the total system size $L$, but it will
depend on the spectrum at length scale $L_{E}$
with number of states $N (E,L_{E})$.
We replace $\xi$ in the above estimation with $L_{E}$,
giving
$$\gamma (E )
\sim
\frac{E^2}{N^2 (E,L_{E})\Delta (L_{E})}
\sim
\frac{E^{3/2}}{E_F^2 \tau^{3/2}} ,$$
in agreement with the prediction of Altshuler and Aronov.\cite{A+A79} 
%
%

\section*{Appendix B}
%
This appendix describes a numerical calculation of the correlation
function in a disk of width $L$ and radius $R$ where $L \ll R$.
The crossover regime between the quasi-0D ($L \ll R \ll L_c$)
and quasi-2D ($L \ll L_c \ll R$) limits is studied.
Using standard diagrammatic perturbation theory techniques,
it is possible to express the correlation function in terms of
diffusion propagators in the disk.\cite{fal97} 
Since $L \ll L_c$, the zero mode dominates the diffusion
propagator in the direction parallel to the current flow, across the width
of the disk, but it is necessary to sum all harmonics of
diffusion perpendicular to the current flow, in the plane of the 
disk.
Adopting circular cylindrical coordinates ${\bf r} = (\rho , \phi ,z)$,
we consider the resonant impurity to be positioned at one side of
the disk $z=0$ and at an arbitrary radius $0 \leq \rho \leq R$ from
the center of the disk.

The numerical procedure outlined here is necessary only in the
crossover regime since the exact geometry of the emitter
is not relevant in the limiting cases:
in the quasi-2D limit a diffusing electron typically does not reach the
boundary of the disk and it is thus possible to integrate over
all harmonics of the diffusion propagator in the plane of the disk
whereas in the quasi-0D limit the zero mode 
in the plane of the disk is not damped very effectively
and it dominates, enabling one to neglect all higher harmonics.
These approximations produce the analytic results given in the main text
and in both cases the position of the resonant impurity $\rho$
is irrelevant.
However the position of the resonant impurity
is crucial in the crossover regime because it is necessary to
sum over many harmonics that are influenced by the exact
geometry of the emitter.

The correlation function of differential conductances can be obtained
from the disordered averaged current-current correlation
function\cite{Lerner} $\langle \delta I(V)\delta I(V')\rangle$
by taking the second derivative with respect to 
$\Delta =\alpha e(V-V')$,
\begin{eqnarray}
\langle \delta G(V)\delta G(V')\rangle 
=-(\alpha e)^2\frac{\partial^2}{\partial\Delta^2}
\langle \delta I(V)\delta I(V')\rangle .
\nonumber
\end{eqnarray}
By expressing the current in terms of Greens functions using
the single-particle Breit-Wigner resonance conductance
formula,\cite{Chaplik,Azbel,ResTunn,LernerRaikh} 
the correlation function takes the form\cite{fal97}
\begin{eqnarray}
\langle \delta G(V)\delta G(V')\rangle
&=& - \frac{1}{\beta}
\left( \frac{G_{\Gamma} \Gamma}{2} \right)^2
\left(\frac{\partial^2}{\partial\Delta^2}\right) 
\nonumber \\
&& \!\!\! \!\!\!
\times
\frac{\Gamma}{\nu}
\int\frac{d\omega \, \left[
P_\omega({\bf r},{\bf r})
+ P_{-\omega}({\bf r},{\bf r})
\right]}
{(\hbar\omega-\Delta)^2+\Gamma^2}
.
\label{p+p}
\end{eqnarray}
In the absence of time-reversal symmetry, the variance changes only by
the standard Dyson's factor of $1/\beta$, where $\beta = 1$ for the
orthogonal ensemble (in the presence of impurity scattering only)
and $\beta = 2$ for the unitary ensemble (in the presence of a finite
magnetic field or weak scattering by magnetic impurities that breaks
time-reversal invariance).
The diffusion propagator in the disk, $P_\omega({\bf r},{\bf r})$,
satisfies the following equation
\begin{equation}
\left[ - D\nabla^2 + \gamma  - i\omega \right]
P_\omega({\bf r},{\bf r}')= \delta ({\bf r}-{\bf r}').
\label{de}
\end{equation}
It describes diffusion in the disk and is therefore
restricted by the tunneling barrier
at $z=0$ and an insulating boundary at the cylinder surface $\rho = R$.
The poorly conducting interface between the emitter and substrate
is also modeled as a tunneling barrier at $z=L$.
These boundary conditions are expressed as
\begin{eqnarray}
\left.\partial_z P\right|_{z=0} = 0, \qquad
\left.\partial_{\rho} P\right|_{\rho = R} = 0, \qquad
\left.\partial_z P\right|_{z=L} = 0 .
\label{bc}
\end{eqnarray}
The correlation function is found by
solving the diffusion equation, Eq.~(\ref{de}),
in the presence of the boundary conditions, Eq.~(\ref{bc}).
The angle $\phi$ is set to zero without loss of generality and we find
\begin{eqnarray}
P_\omega ({\bf r,r}) &=&
\frac{1}{\pi R^2 L}
\sum_{m,\alpha_{nm}}
\frac{J_m^2 \left( \alpha_{nm} \rho /R\right)}
{\left( 1 - m^2/\alpha_{nm}^2 \right) J_m^2 \left( \alpha_{nm} \right)}
\nonumber \\
&& \qquad \times \frac{1}
{\left( D\alpha_{nm}^2/R^2 + \gamma - i\omega \right)} ,
\label{pbes}
\end{eqnarray}
where $m = 0, \pm 1, \pm 2, \ldots$ and
$J_m$ is a Bessel function of the first kind of order $m$.
For a given $m$, the numbers $\alpha_{nm}$ are solutions of the boundary
condition at the cylinder surface $\rho = R$,
\begin{eqnarray}
\left.\partial_{\rho} J_m (\alpha_{nm} \rho /R)
\right|_{\rho = R} = 0 ,
\nonumber
\end{eqnarray}
which may be expressed as
\begin{eqnarray}
m J_m (\alpha_{nm}) = \alpha_{nm} J_{m+1} (\alpha_{nm}) .
\nonumber
\end{eqnarray}
We solve this boundary condition numerically in order
to calculate the propagator, giving the variance and the correlation function
for arbitrary $L_c/R$ and $0 \leq \rho \leq R$.

\begin{figure}
\epsfxsize=\hsize
\epsffile{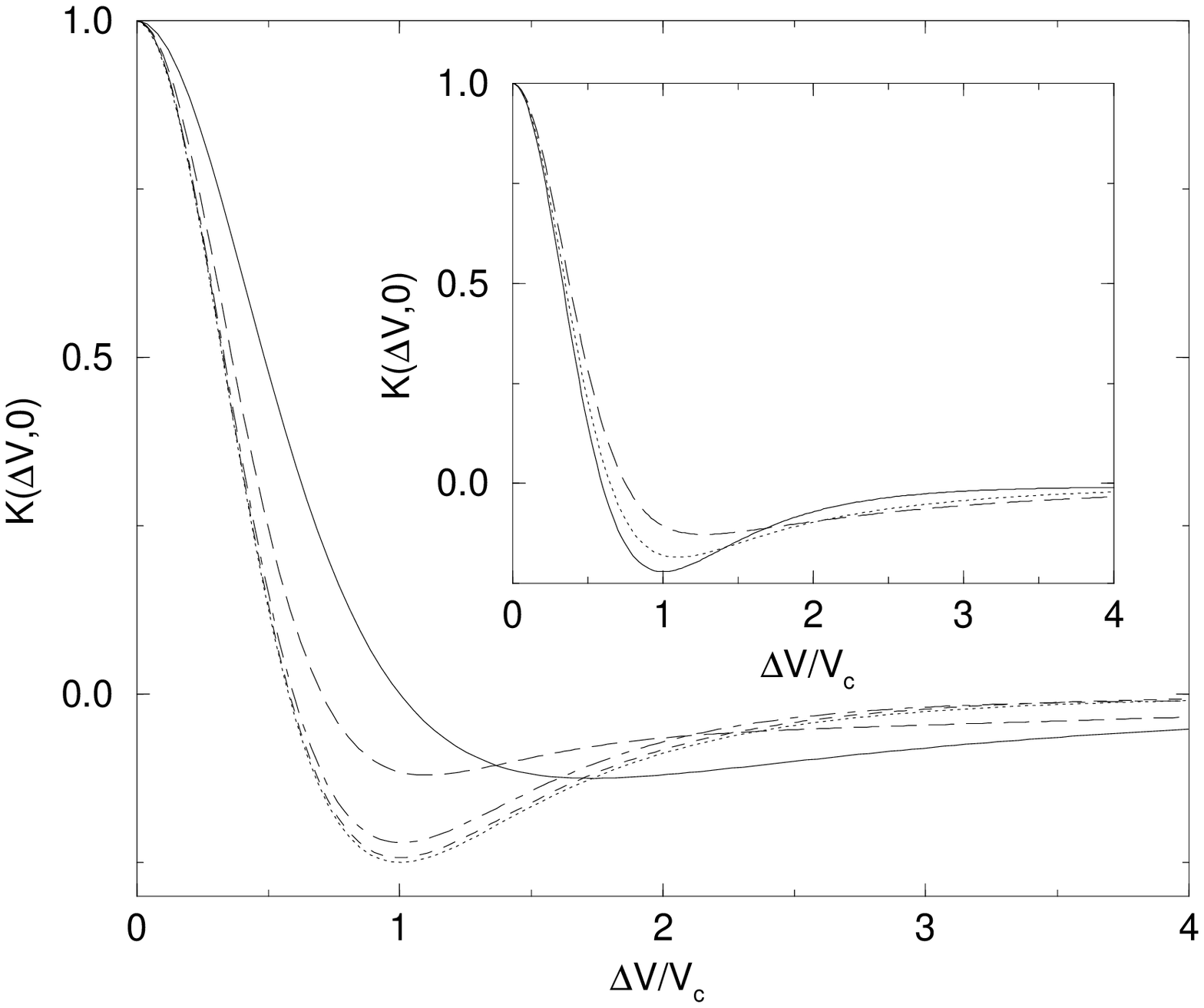}
\refstepcounter{figure}
\label{fig:thfig5}
{\setlength{\baselineskip}{10pt} FIG.\ \ref{fig:thfig5}.
Theoretical form of the correlation function $K(\Delta V)$
as a function of $\Delta V/V_c$
with the resonant impurity at the center of the disk $\rho /R = 0$.
Long dashed line is numerical result for $L_c /R = 1.0$,
dot-dashed for $L_c /R = 1.5$,
and short dashed for $L_c /R = 2.0$.
Solid and dotted lines are the analytic results in 
Eq.~(\ref{cor0d}) for Q0D and Q2D geometry.
Inset is $K(\Delta V,0)$ as a function of $\Delta V/V_c$ for
$L_c /R = 1.5$ and different impurity positions.
Solid line is $\rho /R = 0$, dotted is $\rho /R = 0.5$,
and  long dashed is $\rho /R = 1.0$.} 
\end{figure}

\begin{figure}[t]
\epsfxsize=\hsize
\epsffile{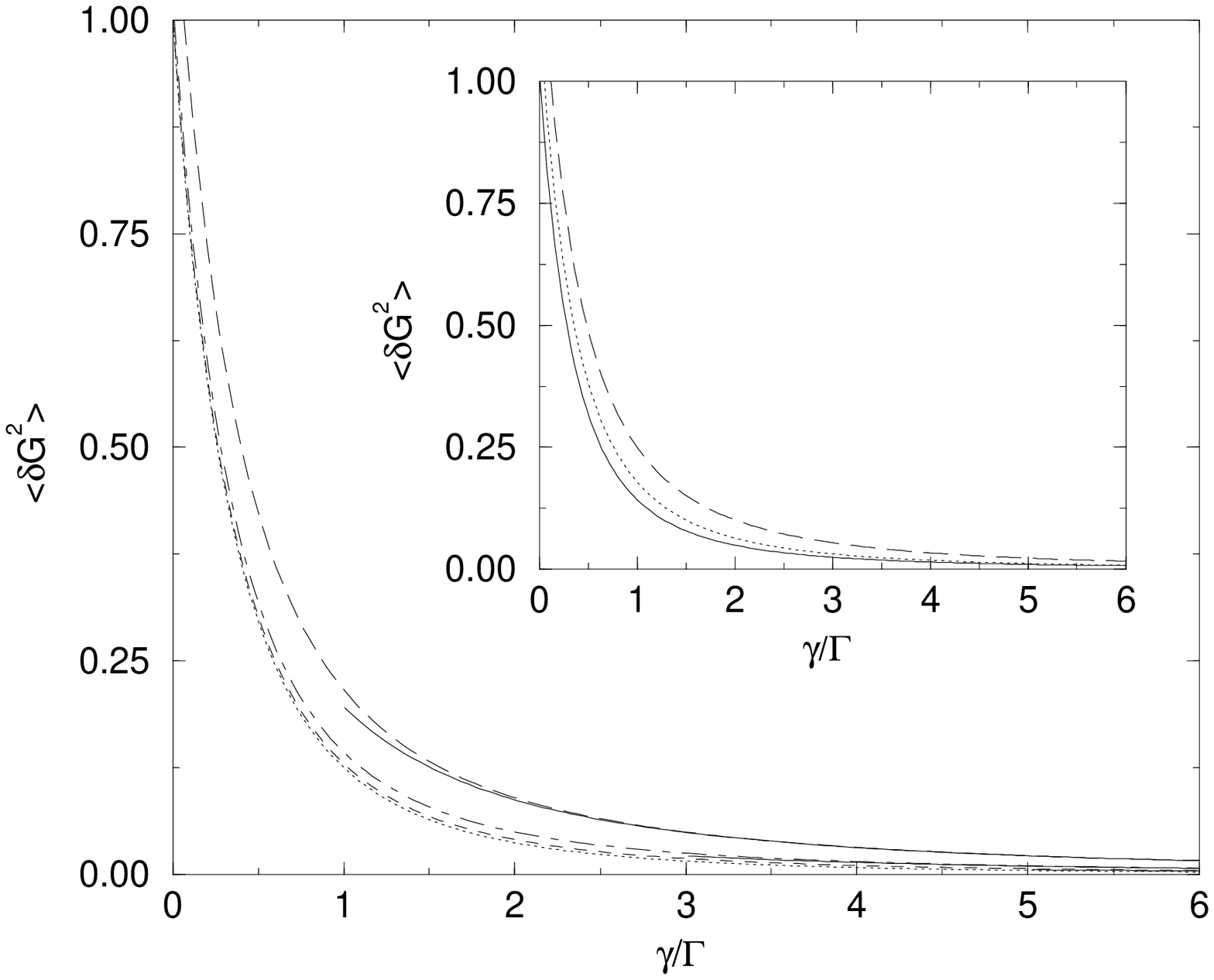}
%
\refstepcounter{figure}
\label{fig:thfig4}
{\setlength{\baselineskip}{10pt} FIG.\ \ref{fig:thfig4}.
Theoretical form of the variance $\langle \delta G^2 \rangle$
as a function of $\hbar\gamma /\Gamma$
with the resonant impurity in the center of the disk $\rho /R = 0$.
From the top, the
long dashed line is the numerical result for $L_{\Gamma} /R = 1.0$,
dot-dashed for $L_{\Gamma} /R = 1.5$,
and short dashed for $L_{\Gamma} /R = 2.0$.
The solid lines show asymptotics at large $\hbar\gamma /\Gamma$
given by the Q2D analytic result Eq.~(\ref{variance})
and the dotted line is the Q0D analytic result Eq.~(\ref{variance}).
All the curves are  normalized by the Q0D analytic result Eq.~(\ref{variance}).
Inset is $\langle \delta G^2 \rangle$ as a function of $\hbar\gamma /\Gamma$
for $L_{\Gamma} /R = 1.5$ and different impurity positions.
Solid line is $\rho /R = 0$, dotted is $\rho /R = 0.5$,
and  long dashed is $\rho /R = 1.0$.} 
\end{figure}

In the main text Fig.~\ref{fig:thfig5} shows the correlation function
for different $L_c/R$ and $\rho /R = 0$ (main part) and for
different impurity positions $\rho /R$ (inset).
We present here for completeness some further numerical results.
Fig.~\ref{fig:thfig4} shows the calculated variance 
$\langle \delta G^2 \rangle$ as a function of $\hbar\gamma /\Gamma$
with the resonant impurity at the center of the disk $\rho /R = 0$
and different values of $L_{\Gamma} /R$.
The short dashed line is the Q0D analytic
result Eq.~(\ref{variance}) whereas the solid lines show asymptotics
at large $\hbar\gamma /\Gamma$ given by the Q2D analytic result
Eq.~(\ref{variance}).
As expected, the numerical plots show behavior similar to the
Q0D analytic form for $L_c \gg R$ (small $\hbar\gamma /\Gamma$)
and similar to Q2D for $L_c \ll R$ (large $\hbar\gamma /\Gamma$)
with a crossover at $L_c \approx R$.
To analyze the effect of the position of the impurity we choose
a particular value of $L_{\Gamma} /R$.
The inset of Fig.~\ref{fig:thfig4} is $\langle \delta G^2 \rangle$
for $L_{\Gamma} /R = 1.5$ and for different impurity positions.
When the impurity position is off-center
the variance has a similar qualitative form as for the impurity on
the cylinder axis, but the fluctuations appear to be generally larger.

\begin{figure}[t]
%
\epsfxsize=\hsize
\epsffile{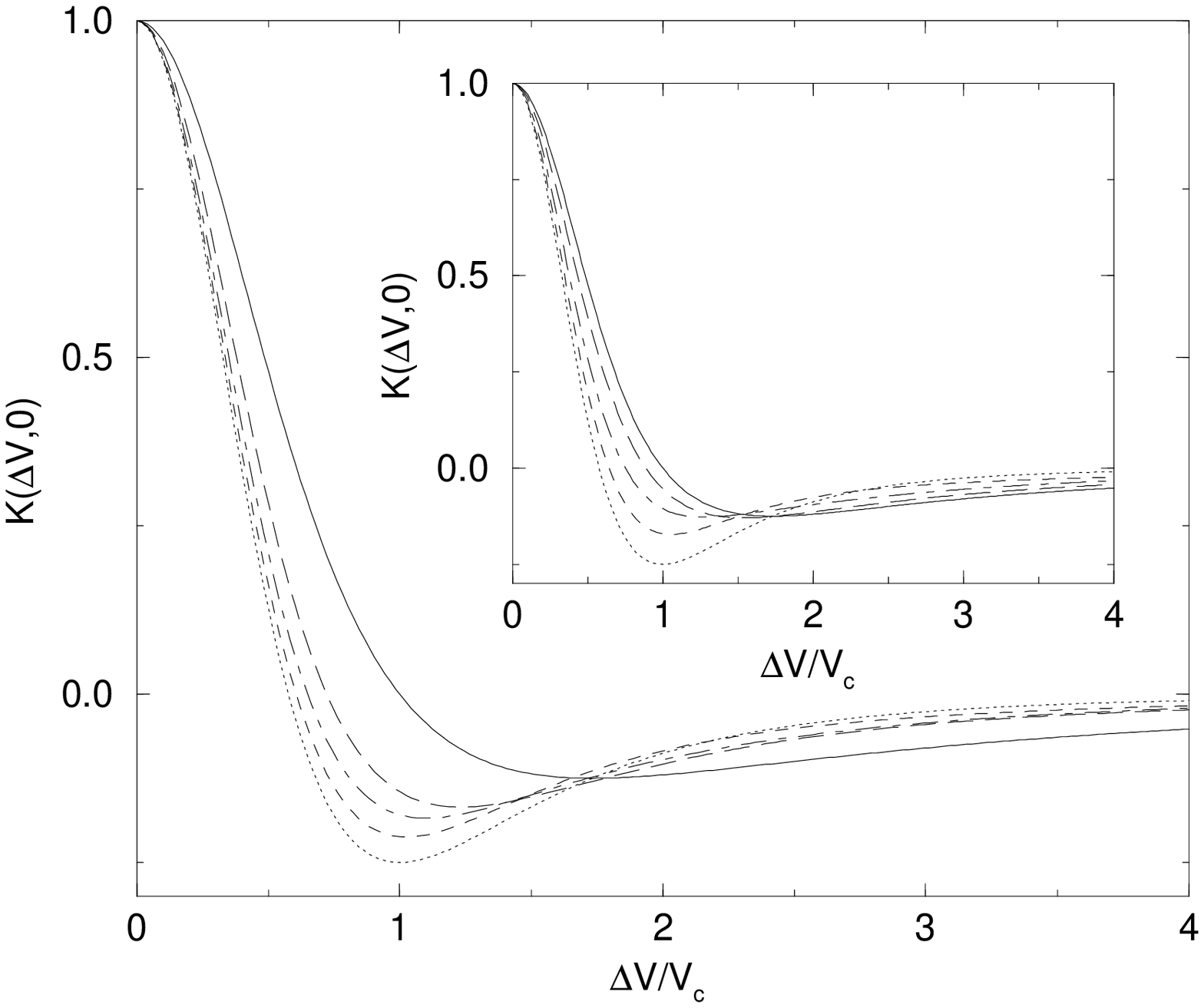}
\refstepcounter{figure}
\label{fig:thfig6}
{\setlength{\baselineskip}{10pt} FIG.\ \ref{fig:thfig6}.
Theoretical form of the correlation function $K(\Delta V,0)$
as a function of $\Delta V/V_c$
with the resonant impurity off center of the disk at $\rho /R = 0.5$.
Long dashed line is numerical result for $L_c /R = 1.0$,
dot-dashed for $L_c /R = 1.5$,
and short dashed for $L_c /R = 2.0$.
Solid and dotted lines are the Q2D and Q0D analytic results 
from Eq.~(\ref{cor0d}).
Inset is $K(\Delta V,0)$ as a function of $\Delta V/V_c$
with the resonant impurity at $\rho /R = 1.0$
with line styles the same as the main part.} 
\end{figure}

Fig.~\ref{fig:thfig6} shows $K(\Delta V,0)$
as a function of $\Delta V /V_c$ and different values of $L_c /R$
for $\rho /R = 0.5$ (main part) and $\rho /R = 1.0$ (inset).
The crossover appears to occur more slowly (over a larger range of $L_c /R$)
when the impurity position is off-center.

\begin{references}
\bibitem{remark1} The quasi-particle to be considered below is, in fact, 
a non-equilibrium hole left after the tunneling process below the Fermi 
level in the emitter. Relaxation of higher-energy electrons into this
emptied state results in a finite lifetime of such hole.
\bibitem{RemSpin}
For a while, we forget about the electron spin,
but the final result will be written in such a form that this 
flaw will be automatically corrected - if there is no 
spin-orbit scattering in the bulk.  Otherwise, the parameter
$\beta$ in the answers should be taken as $\beta=2$.
\end{references}
\end{document}